\documentclass[twocolumn]{aastex631} 
\usepackage{graphicx,apjfonts,amsmath,amssymb,lmodern,graphics}

\newcommand{\rom}[1]{\rm\uppercase\expandafter{\romannumeral #1\relax}}
\usepackage{soul,xcolor}

\accepted{2022/09/19}
\submitjournal{ApJ}

\shorttitle{X-ray obscured AGN J1440}
\shortauthors{Ansh et al.}

\graphicspath{{./}{figures/}}

\begin{document}  

\title{\emph{NuSTAR} observations of a heavily X-ray obscured AGN in the dwarf galaxy J144013+024744}

\correspondingauthor{Shrey Ansh, Chien-Ting Chen}
\email{sa0124@uah.edu, chien-ting.chen@nasa.gov}

\author{Shrey Ansh}
\affiliation{Department of Physics and Astronomy, University of Alabama in Huntsville, 301 Sparkman Dr NW, Huntsville, AL 35899, USA }

\author{Chien-Ting J. Chen}
\affiliation{Science and Technology Institute, Universities Space Research Association, Huntsville, AL 35805, USA}
\affiliation{Astrophysics Office, NASA Marshall Space Flight Center, ST12, Huntsville, AL 35812, USA}
\author{W. N. Brandt}
\affiliation{Department of Astronomy and Astrophysics, 525 Davey Lab, The Pennsylvania State University, University Park, PA 16802, USA}
\affiliation{Institute for Gravitation and the Cosmos, The Pennsylvania State University, University Park, PA 16802, USA}
\affiliation{Department of Physics, The Pennsylvania State University, University Park, PA 16802, USA}
\author{Carol E. Hood}
\affiliation{Department of Physics, California State University, San Bernardino, 5500 University Parkway, San Bernardino, CA 92407, USA}

\author{E. S. Kammoun}
\affiliation{IRAP, Université de Toulouse, CNRS, UPS, CNES 9, Avenue du Colonel Roche, BP 44346, F-31028, Toulouse Cedex 4, France}
\affiliation{INAF – Osservatorio Astrofisico di Arcetri, Largo Enrico Fermi 5, I-50125 Firenze, Italy}
\author{G. Lansbury}
\affiliation{European Southern Observatory, Karl-Schwarzschild- Strasse 2, D-85748 Garching, Germany}
\author{Stéphane Paltani}
\affiliation{Department of Astronomy, University of Geneva, 1290 Versoix, Switzerland}
\author{Amy E. Reines}
\affiliation{eXtreme Gravity Institute, Department of Physics, Montana State University, Bozeman, MT 59717, USA}
\author{C. Ricci}
\affiliation{Núcleo de Astronomía de la Facultad de Ingeniería, Universidad Diego Portales, Av. Ejército Libertador 441, Santiago, Chile}
\affiliation{Kavli Institute for Astronomy and Astrophysics, Peking University, Beijing 100871, People’s Republic of China}
\author{Douglas A. Swartz}
\affiliation{Science and Technology Institute, Universities Space Research Association, Huntsville, AL 35805, USA}
\affiliation{Astrophysics Office, NASA Marshall Space Flight Center, ST12, Huntsville, AL 35812, USA}
\author{ Jonathan R. Trump}
\affiliation{Department of Physics, 196A Auditorium Road Unit 3046, University of Connecticut, Storrs, CT 06269 USA}

\author{F. Vito}
\affiliation{INAF – Osservatorio di Astrofisica e Scienza dello Spazio di Bologna, Via Gobetti 93/3, I-40129 Bologna, Italy}
\author{Ryan C. Hickox}
\affiliation{Department of Physics and Astronomy, Dartmouth College, 6127 Wilder Laboratory, Hanover, NH 03755, USA}

\setstcolor{red}

\begin{abstract}
We present a multi-wavelength analysis of the dwarf Seyfert-2 galaxy J$144013+024744$, a candidate obscured active galactic nucleus (AGN) thought to be powered by an intermediate-mass black hole (IMBH, $M_\bullet \approx 10^{4-6} M_\odot$) of mass $M_{\bullet} \sim 10^{5.2}M_\odot$. To study its X-ray properties, we targeted J$144013+024744$ with NuSTAR for $\approx 100~\rm ks$. The X-ray spectrum was fitted with absorbed power law, Pexmon and a physical model (RXTorus). A Bayesian X-ray analysis was performed to estimate the posteriors. The phenomenological and the physical models suggest the AGN to be heavily obscured by a column density of $N_{\rm H} = (3.4-7.0)\times10^{23}$ cm$^{-2}$. In particular, the RXTorus model with a sub-solar metallicity suggests the obscuring column to be almost Compton-thick. 
We compared the 2--10~keV intrinsic X-ray luminosity with the inferred X-ray luminosities based on empirical scaling relations for unobscured AGNs using $L_{[O\rom {4}](25.89\rm \mu m)}$, $L_{[O\rom {3}](5007)\text{{\AA}}}$, and $L_{6\rm \mu m}$ and found that the high-excitation $[O\rom{4}]$ line provides a better estimate of the intrinsic 2--10~keV X-ray luminosity ($L_{2-10}^{\rm int} \sim 10^{41.41}$~erg~s$^{-1}$)).
Our results suggest that J$144013+024744$ is the first type-2 dwarf galaxy that shows X-ray spectroscopic evidence for obscuration. The column density that we estimated is among the highest measured to date for IMBH-powered AGNs, implying that a typical AGN torus geometry might extend to the low-mass end. This work has implications for constraining the black hole occupation fraction in dwarf galaxies using X-ray observations.
\end{abstract}
\keywords{
 galaxies: dwarf  ---  galaxies: active}
\section{Introduction}\label{sec:intro}
There is a general consensus that Active Galactic Nuclei (AGNs) are powered by the accretion of material onto a supermassive black hole (SMBH, $M_\bullet > 10^6 M_\odot$), but the origin of these SMBHs remains an open question. One of the more quantifiable constraints to differentiate SMBH seeding scenarios is to study the BH occupation fraction in local dwarf galaxies, as they are considered to have undergone fewer mergers and therefore contain the ``fossil records'' of the first SMBHs (e.g.\citealt{volonteri2010formation}). The first step toward understanding the BH occupation fraction is to understand the full picture of the dwarf galaxy population with an actively accreting nucleus \citep[for a review, see][]{Reines2022}. However, measuring the active fraction is already challenging for these less-luminous AGNs that are powered by less-massive SMBHs or even intermediate-mass black holes (IMBHs, $M_\bullet\approx 10^{4-6} M_\odot$), and obscuration further exacerbates the problem as typical AGN signatures in soft X-rays and UV-optical bands become almost indiscernible from the stellar emission from their host galaxies
\citep[e.g.,][]{grimm2003high,mineo2012x,trump2015biases}.

\citet[R13 onwards]{reines2013dwarf} identified 136 optically selected AGNs out of $\sim 25,000$ dwarf emission-line galaxies using SDSS spectroscopy, finding an active fraction of $\sim 0.5$\%. Similarly, \citet{baldassare2018identifying} used $\sim 28,000 $ galaxies from SDSS Stripe 82 to search for AGN activity based on optical variability and found that the active fraction decreases with stellar mass and is significantly lower for  galaxies with masses $ < 10 ^{10} M_{\odot}$.
However, measurements from optical surveys are limited as X-ray and mid-IR surveys have demonstrated that for massive galaxies, optical surveys can miss up to $50 \%$ of the AGN population (see \citealt{hickox2018obscured} for a detailed review). While space-based X-ray and mid-IR observatories have provided a more complete view of the AGN census in massive galaxies, the obscured AGN population in dwarf galaxies remains extremely elusive. This largely owes to the fact that the luminosity of AGNs in dwarf galaxies can be orders of magnitude lower than that of AGNs in regular galaxies, making typical AGN identifiers such as optical emission-line ratios and mid-IR colors easily buried by the host-galaxy light \citep{trump2015biases,Hainline2016}. Moreover, mid-IR searches for AGNs in dwarf galaxies suffer from severe contamination \citep[][]{Satyapal2014,Kaviraj2019}, as dwarf starburst galaxies can mimic the mid-IR colors of more luminous AGNs \citep[][]{Hainline2016,Latimer2021_wise}.
On the other hand, X-ray observations are less susceptible to contamination from the host-galaxy and are a preferred method to detect AGNs in dwarf galaxies and measure their obscuration level \citep{lemons2015x, xue2012tracking, brandt2015cosmic,Birchall2020,Latimer2021_erosita}. 
In particular, hard X-rays ($>10$ keV) are extremely useful 
due to their strong penetrating power that can even overcome the heavy obscuration of a torus \citep{mushotzky2004agn}

For AGNs hosted by dwarf galaxies, 
studies employing X-ray spectral analysis to constrain the properties of the obscuring material 
remain scarce due to limited source counts. For the few studies that attempted to do so \citep{dong2012x,ludlam2015x,baldassare2017x}, 
the X-ray spectra of these optically-selected broad-line AGNs were largely explained by a simple absorbed power law with little to no obscuration. 
An exception is NGC~4395. This nearby archetypal IMBH-powered type 1 AGN was found to have highly variable fluxes with evidence for varying partial-covering neutral
absorption with a moderate column density of $10^{22-23}$~cm$^{-2}$ \citep{moran2005extreme,parker2015revealing,kammoun2019nature}.
However, the existence of more heavily obscured (i.e., $N_{\rm H}>10^{23}$~cm$^{-2}$) AGN in local dwarf galaxies remains an open question, although high redshift X-ray stacking analyses of similar galaxies do suggest this population exists  \citep[e.g.,][]{xue2012tracking,mezcua2016population}.
Therefore, the search for heavily obscured AGNs in dwarf galaxies remains a necessary step for reliably using these systems as a constraint for primordial BH seeding scenarios.

\begin{figure}
    \centering
    \includegraphics[width=0.5\textwidth]{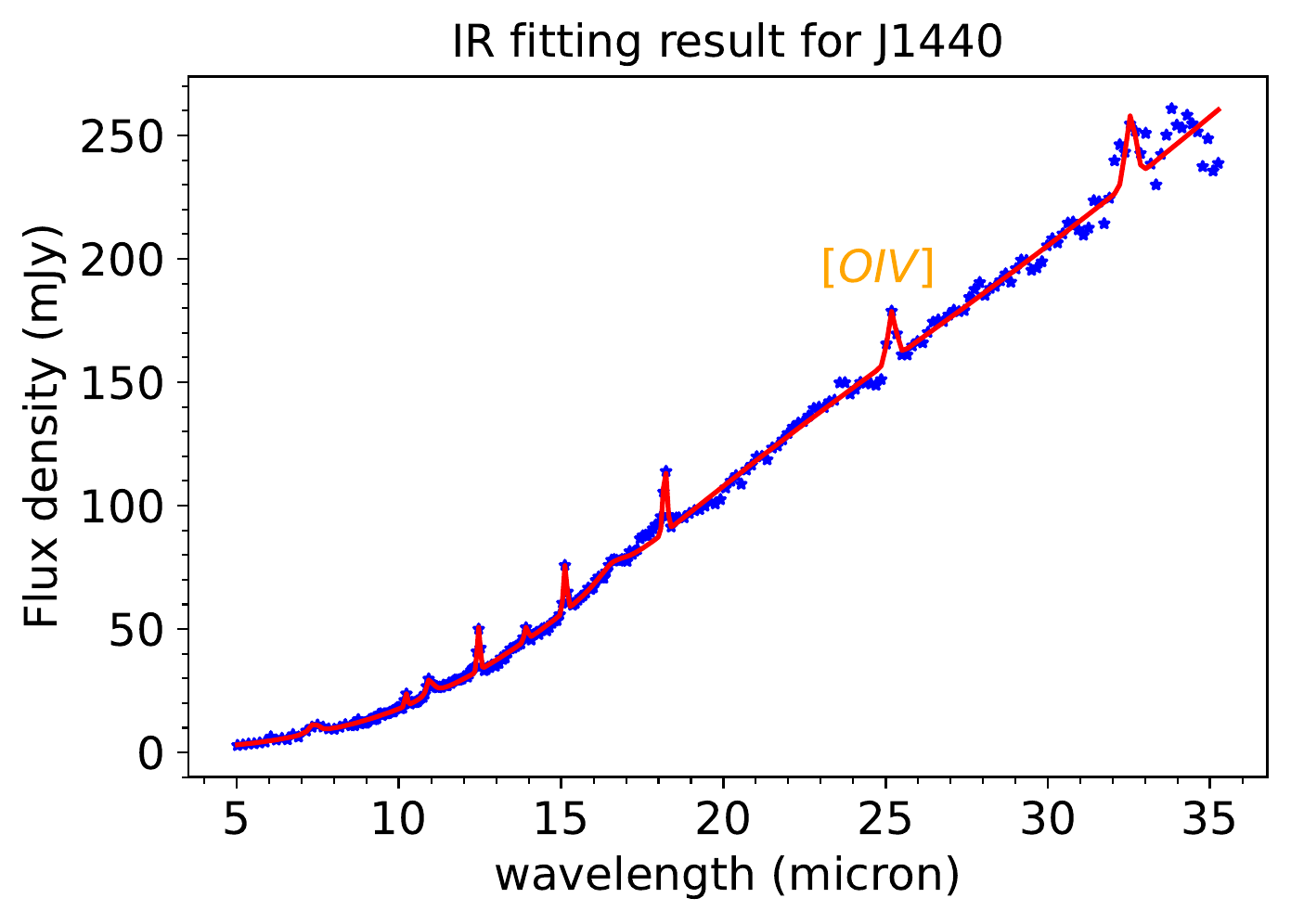}
    \caption{
            \footnotesize
Spitzer IRS spectrum of J1440. The blue stars show the IR spectral data obtained from Spitzer IRS and the red line shows the model obtained by H17.
    The presence of the $[O\rom {4} ]$ $25.89~\rm \mu m$ emission-line (labeled in the figure) is consistent with AGN activity. } 
    \label{fig:irsspectrum}
\end{figure}

This paper focuses on a promising obscured AGN candidate hosted by the dwarf galaxy J$144012.70+024743.5$\footnote{also known as Tol 1437+030 \citep{bohuski1978nature}} \footnote{"RCG 32" in R13 sample of dwarf galaxies} (J1440 onward). J1440 is a dwarf galaxy located at $\rm z=0.029$ and was selected from the Seyfert 2 low-mass galaxy sample with very low velocity dispersion based on the Sloan Digital Sky Survey (SDSS) data by \citet{barth2008low}. Despite its Seyfert 2 classification, R13 identified this target to have a weak broad $\rm H\alpha$ component in its optical spectrum.
The AGN activity can also be confirmed based on WISE selection criteria which gives a magnitude difference between W1 and W2 as 1.14, well above the required difference of 0.77 \citep{assef2018wise}. \citet{thornton2009emission} used a 22.92~ks exposure with XMM-Newton to study this target (though 25.3$\%$ of the exposure time was lost due to background flaring). The signal-to-noise ratio was too low at $>1$~keV to make any definite conclusion about the column density.
Another observation of this target was made in 2015 using Chandra but the source was not detected in the hard band, and the total counts were too low to perform a spectral analysis \citep{baldassare2017x}.
\citet[H17 onwards]{hood2017spitzer} followed up the \cite{barth2008low} sample with the Spitzer IRS survey, and found some objects in this sample with substantial mid-IR coronal lines such as $[Ne \rom {5} ]$(14.32$~\rm \mu m$) and $[ O \rom {4} ]$(25.89$~\rm \mu m$). 
Encouraged by the presence of weak X-ray emission and the high-excitation $[ O \rom {4} ]$(25.89$~\rm \mu m$, see Figure~\ref{fig:irsspectrum}) 
emission-line, an AGN marker which is less affected by the host-galaxy contamination, we targeted J1440 with the Nuclear Spectroscopic Telescope Array \citep[NuSTAR,][]{nustar2013} for 100~ks. Section 2 describes the AGN and galaxy properties of J1440 by fitting the optical-IR photometric data with CIGALE \citep[Code Investigating GALaxy Emission,][]{boquien2019cigale}. In Section 3, we describe the X-ray spectral models used to fit the NuSTAR data and the results of our X-ray spectral analysis. We supplement our X-ray analysis with multi-wavelength data available for J1440 in Section 4. We present our conclusions and discussions in Section 5. In this paper, we used the standard $\Lambda$CDM cosmology with H$_o$ = 70.4 km s$^{-1}$ Mpc$^{-1}$ and $\Omega_{\lambda} = 0.73$.

 \begin{figure}
    \centering
    \DeclareGraphicsExtensions{png}
    \hspace*{-0.2cm}
    \includegraphics[width=0.5\textwidth]{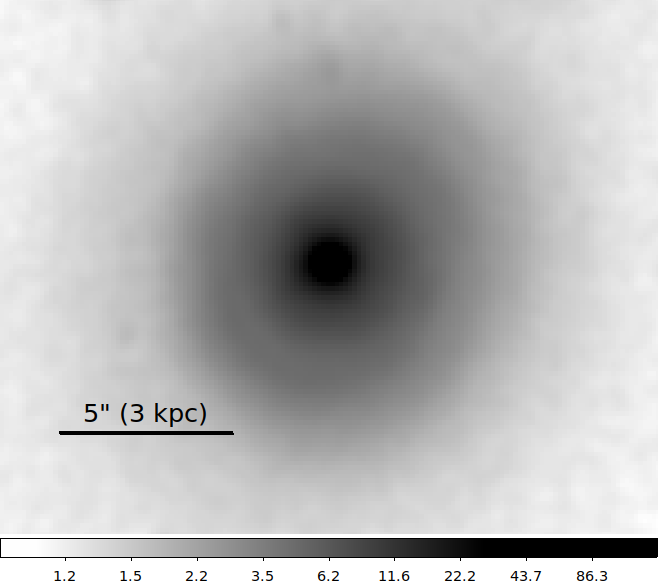}
    \caption{
    \footnotesize
    Hubble Space Telescope WFC3/F110W near-IR image for J1440. The galaxy was classified as a type Sa galaxy in  \citet{Schutte2019} with a clear bulge component at its core. 
    }
    \label{fig:hst}
\end{figure}

 \begin{figure*}
    \centering
    \includegraphics[width=\textwidth]{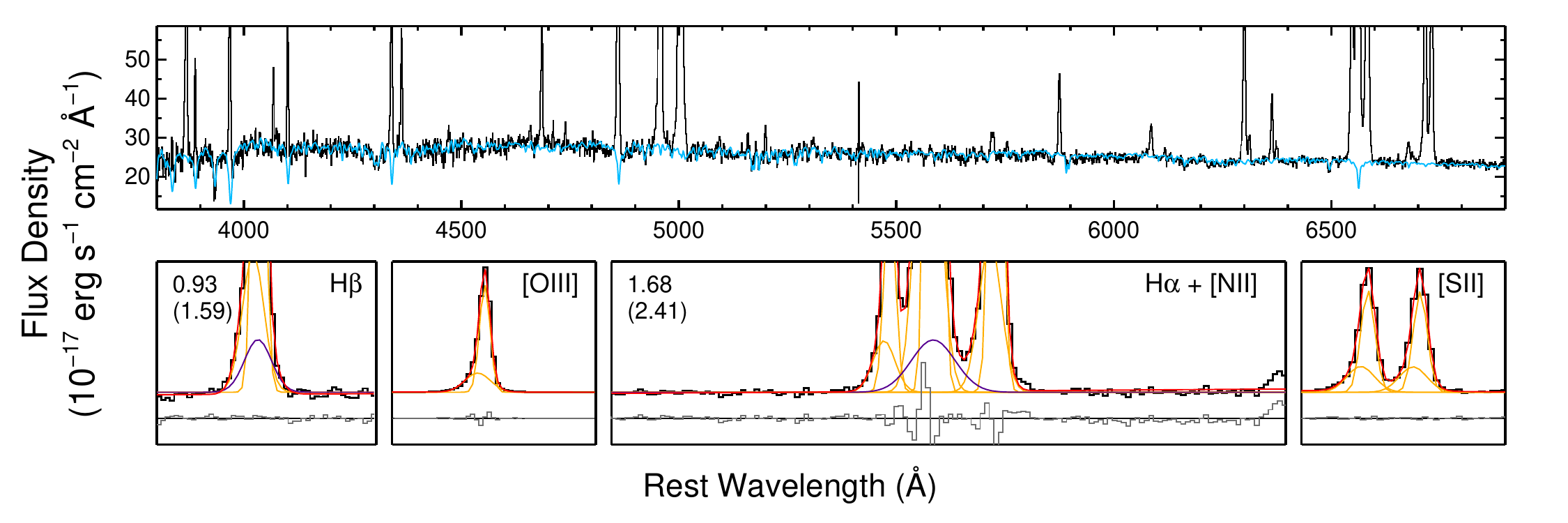}
    \caption{ Optical spectrum for J1440 with zoomed-in line profile fits of key emission-lines as analyzed in R13. The broad components in $\rm H_{\rm\alpha}$ and $\rm H_{\rm\beta}$, as well as the strong $[O\rom{3}](5007$\text{\AA}) and $[S\rom{2}]$ lines are indicative of AGN activities.} 
    \label{fig:opticalspectrum}
\end{figure*}

\begin{figure*}
     \centering
     \plottwo{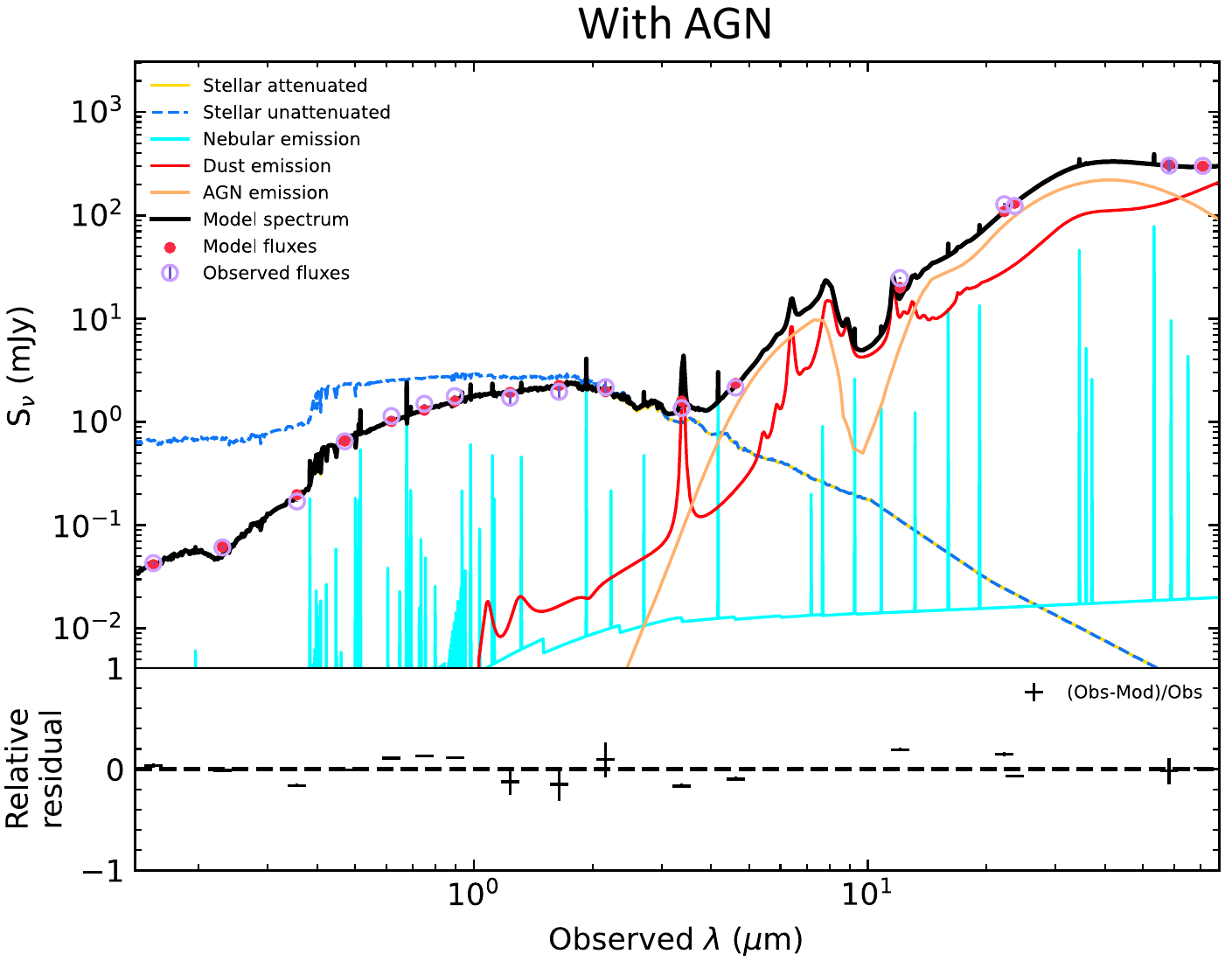}{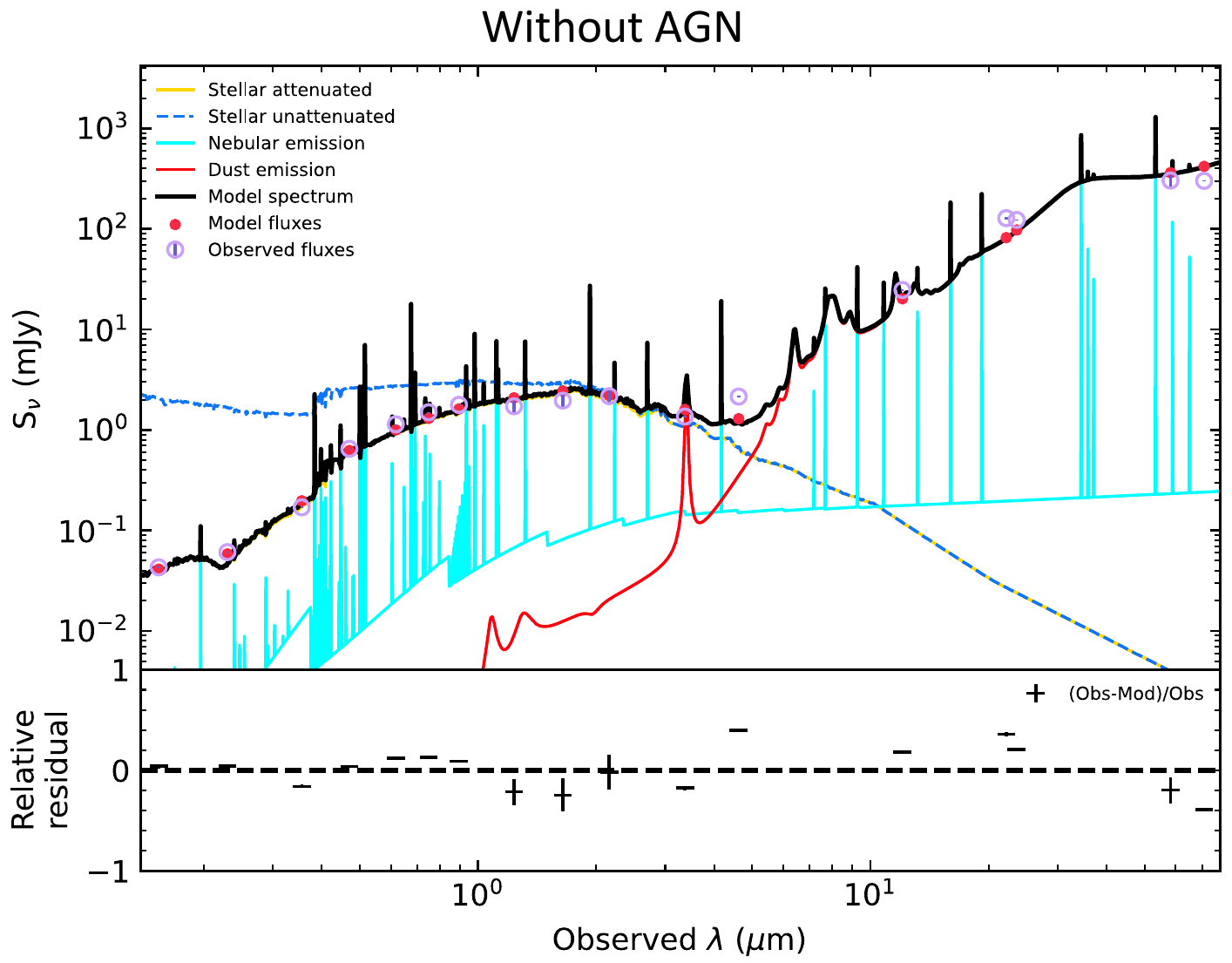}

        \caption{
        \footnotesize
        SED (Spectral energy distribution) fitting of Optical-IR photometric data of J1440 using the CIGALE software. The plot with AGN (left) and without AGN (right) are shown with its residual. The higher residual in the right plot suggests the need for an AGN component. Moreover, the AGN component in the optical band is heavily suppressed which suggests the presence of an obscuring medium.  }
        \label{fig:SED}

\end{figure*}

\section{Multi-wavelength Data for J144013+024744  } \label{sec:data}

\subsection{Black hole mass estimates}
There are multiple ways to estimate the mass of the central BH in galaxies. 
For instance, bulge-dominated galaxies are expected
to trace the $M_\bullet-\sigma_\star$ relation to an acceptable accuracy. 
\citet{Schutte2019} studied J1440 using deep Hubble Space Telescope/WFC3 observations and found that J1440 is an Sa galaxy with dim spiral regions in the disc (see Fig.~\ref{fig:hst}). J1440 was found to have a classical bulge with the bulge Sérsic index \citep{sersic1968atlas} of $n=1.6$ while the Sérsic index for the disc component was $n=0.75$. 
Supported by the presence of a prominent bulge component, 
we utilize the $M_{\bullet}-\sigma_{\star}$ relation derived for low mass BHs by \citet{xiao2011exploring}; $ \log( M_{\bullet}) = (7.68 \pm 0.08) + ( 3.32 \pm 0.22) \log(\sigma_{\star}/200$ $\rm km~s^{-1})$.
For J1440, this leads to an estimated BH mass of $M_{\bullet} = 10^{5.5\pm 0.2}M_\odot$ \citep[based on $\sigma_\star = 44\pm4$ km~s$^{-1}$ from][]{barth2008low}. 

The BH mass can also be estimated from the velocity dispersion of the AGN broad-line region using single-epoch spectra, which is done by assuming the broad-line region gas to be virialized and 
and follow empirical BH radius-luminosity relations. 
R13 used the broad ${\rm H}_{\alpha}$ line to obtain a viral BH mass estimate of $M_\bullet = 10^{5.2}M_\odot$ (see Figure~\ref{fig:opticalspectrum} for the fitting of the optical spectrum). 

The two $M_\bullet$ measurements discussed here are consistent with each other within their respective uncertainties, and both are in the range of what is conventionally considered as an IMBH ($<10^6M_\odot$).

\subsection {SED fitting using UV-Optical-IR photometric data}\label{sec:sed}
In this section, we utilize the existing photometric data of J1440\footnote{We collected photometric measurements from the \citet{https://doi.org/10.26132/ned1}} to estimate the physical properties of the galaxy such as its stellar mass, luminosity, star formation rate (SFR) and metallicity using CIGALE. The photometric data we used are: Far and Near-UV from GALEX (Kron flux density in an elliptical aperture), Optical from SDSS (SDSS model C), Near-IR from 2MASS (2MASS XSC), and IR from WISE (Profile-fit), Spitzer-MIPS (PSF flux density) and IRAS (fixed aperture). 
These photometric data are initially fitted with the modules accounting for the host-galaxy emission only. 
The best-fit SED deviates from the data significantly at the IR wavelengths,
with a poor fit statistic of $\chi_{reduced}^2 =4.06$. This suggests the galaxy-only modules are not sufficient for J1440. 
We then include an AGN component, \texttt{Fritz2006}, 
which is an AGN template library comprised of an isotropic point-source emission component and a thermal and scattering dust torus emission component. 
\cite[see][for details]{fritz2006revisiting,boquien2019cigale}. This 
greatly reduced the disagreement between the data and best-fit SED in the IR wavelengths (see Figure \ref{fig:SED}), with an improved fit statistic of $\chi_{reduced}^2 =1.12$ (for $\rm Z=0.004$\footnote{Here Z is defined as the mass-fraction of elements heavier than helium.}, we discuss the fit statistics for different values of Z later in this section). To confirm the need for an AGN component, we utilized the Akaike information criterion \citep[AIC,][]{akaike1974new}. AIC can be expressed as AIC = $\chi^2 +2k $, here $\chi^2$ determines the goodness of fit and was 17.95 for the model including an AGN component and 64.97 for the model with no AGN (galaxy only). AIC penalizes for extra degrees of freedom through $k$. As we are interested in $\Delta AIC$, the value of $\Delta k$ on adding the AGN component was 7. A significant result ($3 \sigma $) is obtained when AIC changes by 7 (see \citealt{yang2018does} and references therein). For J1440, $\Delta \rm AIC$=$\rm AIC_{\rm no~AGN}$-$\rm AIC_{\rm AGN}$ was 33.02, which was highly significant and confirmed that an AGN component is required to explain the SED of J1440 (the lower the value of AIC, better the fit).

For our SED fitting involving both the galactic and AGN components, we explored three different values of metallicity; $\rm Z=0.0004$, $0.004$ and $0.02$\footnote{0.02 being the solar metallicity}. The $\chi_{reduced}^2$ for these three metallicities are 0.91, 1.12 and 1.6 respectively.
We note that the measurement of metallicities based on photometric SED fits might have limited accuracy, and the spectroscopic measurements can also be challenging due to the presence of an AGN. However, galaxies with mass similar to J1440 are expected to have lower metallicities than their more massive counterparts, as suggested by the mass-metallicity relation \citep[e.g.,][]{tremonti2004origin,gao2018mass}. 
Since the reduced $\chi^2$ for $\rm Z=0.0004$ and $\rm Z=0.004$ are both acceptable, we consider the recent study by
\citet{ma2016origin}, who used high-resolution cosmological zoom-in simulations to study the mass-metallicity relation over a wide range of stellar mass and redshift. Using their redshift-dependent mass-metallicity relation and the best-fit $M_{\star}$ for J1440, the expected metallicity for J1440 is $\rm Z=0.0039$, which is consistent with our choice of metallicity in the CIGALE SED fitting, and implies J1440 to be a galaxy with sub-solar metallicity. The galactic and AGN properties estimated from the CIGALE SED fitting (for $\rm Z=0.004$) are:
\begin{enumerate}
  \item $M_{\star} = 10^{9.20 \pm 0.05}M_\odot$ (This stellar mass obtained by CIGALE is slightly lower than that obtained by R13 ($M_{\star}= 10^{9.40}M_{\odot}$) using kcorrect code that fits broadband fluxes using stellar population synthesis models).
  \item $\rm SFR =  0.42 \pm 0.24 $ $\rm M_{\odot}~\rm yr^{-1}$. The SFR and $M_{\star}$ values place J1440 on the star-forming main sequence based on, e.g., \cite{whitaker2012star}.
  \item $L_{\rm AGN}= (4.33 \pm 0.35)\times10^{43}$ $\rm ergs~s^{-1} $(AGN bolometric luminosity).
\end{enumerate}  
We list the input parameter ranges and the best-fit values for all the CIGALE modules in Table~\ref{tab:cigaleresult}.

\begin{table*}
\centering
\caption{\footnotesize
CIGALE templates and best fit values:
The modules that we used in the fitting process are:(a) [sfh2exp] Models the SFR in terms of decreasing exponential (b) [bc03]: Models the intrinsic stellar spectrum (c) [nebular]: Calculates the nebular emission-lines (d) [dustatt$\_$calzleit]: Models the attenuation due to the dust (e) [dl2007]: Calculates reprocessed UV to NIR emissions due to dust (f) [fritz2006]: Models the AGN emission (g) redshifting}
\footnotesize
\begin{tabular}{ccc}
\hline
\hline
Module\footnote{More details on the individual modules can be found at https://cigale.lam.fr/documentation/}  &  Initial values & Best fit\\
(a) \texttt{sfh2exp} & & \\
\hline
$\tau_{\rm main}$ & $100, 300, 1000, 3000, 10000$ & 300  \\
tau$\_$burst & 10, 50 & 10 \\
f$\_$burst & 0.01 , 0.1 & 0.1 \\
age &  100, 500, 1000, 2000, 5000 , 10000 & 1000 \\
burst$\_$age & 10, 25, 50, 100 & 100 \\
\hline
(b) \texttt{bc03 (Chabrier IMF)} & & \\
\hline
 metallicity & 0.004, 0.004, 0.02 & \\
 separation$\_$age & 10 & \\
 \hline
 (c) \texttt{ nebular} & & \\
 \hline
 logU & -3.0 & \\
 f$\_$esc & 0.0 & \\
 f$\_$dust & 0.0 & \\
 lines$\_$width & 300.0 & \\
 \hline
(d) \texttt{[dustatt$\_$calzleit]} & &\\
\hline
 E$\_$BVs$\_$young  &  0.2, 0.4, 0.6, 0.8  & 0.6 \\ 
E$\_$BVs$\_$old$\_$factor & 0.44 & \\
uv$\_$bump$\_$wavelength & 217.5 & \\
uv$\_$bump$\_$width & 35.0 &\\
uv$\_$bump$\_$amplitude &  1.0 , 2.0 & 2.0 \\ 
powerlaw$\_$slope & 0.0 & \\
\hline
(e) \texttt{ dl2007} & &  \\
\hline
 qpah &  0.47, 1.77, 2.50 & 1.77 \\
 umin & 0.10, 0.50, 1.0 & 1.0 \\
 umax & 1000000.0 & \\
 gamma & 0.1 &  \\
 \hline
(f) \texttt{Fritz2006} & &  \\
 \hline
r$\_$ratio &  60.0 &  \\
tau & 3.0, 6.0, 10.0 & 10.0 \\
beta & -1.00, -0.50, 0.00 & -0.50 \\
gamma & 0.0 & \\
opening$\_$angle &  140 & \\
psy & 20.100 , 40.100, 60.100 & 60.100 \\
fracAGN & 0.1 , 0.3 , 0.5  & 0.5  \\

\hline \hline
\label{tab:cigaleresult}
\end{tabular}
\end{table*}

We also compared the SFR obtained from CIGALE with that obtained from the polycylic aromatic hydrocarbon (PAH) emission \citep{o2009polycyclic}. These complex molecules often break apart when subjected to high-energy photons from the AGN \citep{schweitzer2006spitzer}, therefore, their presence can be used to determine the SFR and the temperature of the dust since these molecules can only form in the colder region of the galaxy (see H17 and references therein). H17 estimated the SFR for J1440 from the PAH features at $7.7 \rm \mu m$ and $11.3 \rm \mu m$. The SFR value of $0.16\pm 0.10 $ $\rm M_{\odot}~yr^{-1}$ was lower but within the uncertainty range of SFR estimated from CIGALE. 
\subsection{Existing soft X-ray data from XMM-Newton and Chandra}\label{sec:xmm}
J1440 was targeted by XMM-Newton in 2006  (XMM-Newton ObsID$=$0400570101) with an exposure of 22.92~ks \citep{thornton2009emission}, and was also targeted by Chandra in 2015 for $\approx 6$~ks (Chandra ObsID$=$17035).
The XMM-Newton data was found to be heavily background dominated above $1~\rm keV$ as suggested by the 4XMM catalog, which is based on the XMM data analysis pipeline provided by the XMMSOC \citep{webb2020xmm}. The target was strongly detected in the $<1$~keV bands with ${\rm DET\_{ML}} > 150$\footnote{${\rm DET\_{ML}}$ is the detection maximum likelihood computed by the XMM-Newton source detection, {\sc emldetect}, see \url{https://xmm-tools.cosmos.esa.int/external/sas/current/doc/emldetect.pdf}.}. However, for harder bands the detection likelihood rapidly drops to $\sim 26$ for 1--2~keV to $\sim 8$ and $\sim 5$ for 2--4.5~keV and 4.5--12~keV bands, respectively. 
While the source is likely to be detected in the 1--2~keV band according to the 4XMM pipline software, there are only $\sim 3\pm6$ counts in the 1--2 keV band within our spectral extraction region, essentially rendering spectral fitting infeasible beyond 1~keV. 
Note the ${\rm DET\_{ML}}$ values represent the likelihood for the source to be detected according to the algorithm of {\sc emldetect}, and XMMSOC suggests a minimum of ${\rm DET\_{ML}}=10$ for a reliable detection\footnote{\url{https://www.cosmos.esa.int/web/xmm-newton/sas-thread-src-find-stepbystep}.}.
Monte Carlo simulations also suggest the true reliability of a given ${\rm DET\_{ML}}$ value can be non-trivial.  For instance, studies in deep XMM-Newton surveys suggest that the ${\rm DET\_{ML}}$ values required for a hard ($2-12$~keV) band source to be detected with a 99.7\% ($3\sigma$) reliability is $\approx 10$ \citep{chen2018xmm} or higher \citep{ni2021xmm}.
Due to the background-dominated spectrum for this target, \cite{thornton2009emission} was only able to analyze the $<1$~keV spectrum, where they found the object to be dominated by a diffuse plasma model \citep[see][for details]{thornton2009emission}.

The Chandra data were also heavily dominated by the soft X-ray emission with only one $>2$~keV count \citep[see Table 2 in][]{baldassare2017x}, which is effectively a non-detection. Due to the low counts in the Chandra data, no spectral analysis was done in \cite{baldassare2017x}. They also found that the soft X-ray flux for this object is consistent with the expected value derived based on the scaling relation between star formation rate and X-ray luminosity for high-mass X-ray binaries (see their Sec. 4.1 for details).

Both XMM-Newton and Chandra data suggest the soft X-ray component for J1440 is dominated by the X-ray emission not form the X-ray corona near the accretion black hole, but rather is associated with the host galaxy stellar activity or the diffuse emission in the narrow-line regions \citep[e.g.,][]{bianchi2006soft, bogdan2017}. However, 
\citet{thornton2009emission} suggested the discrepancy between the soft X-ray emission and the $L_{[O\rom{3}](5007 \rm \text{\AA})}$-$L_{\rm 2-10 keV}$ relation established by \citet{panessa2006x} may imply some level of X-ray obscuration for J1440. This is investigated with NuSTAR observations discussed in the subsequent sections.

\subsection{NuSTAR data}
\begin{figure}
    \DeclareGraphicsExtensions{.png}
    \includegraphics[width=\columnwidth]{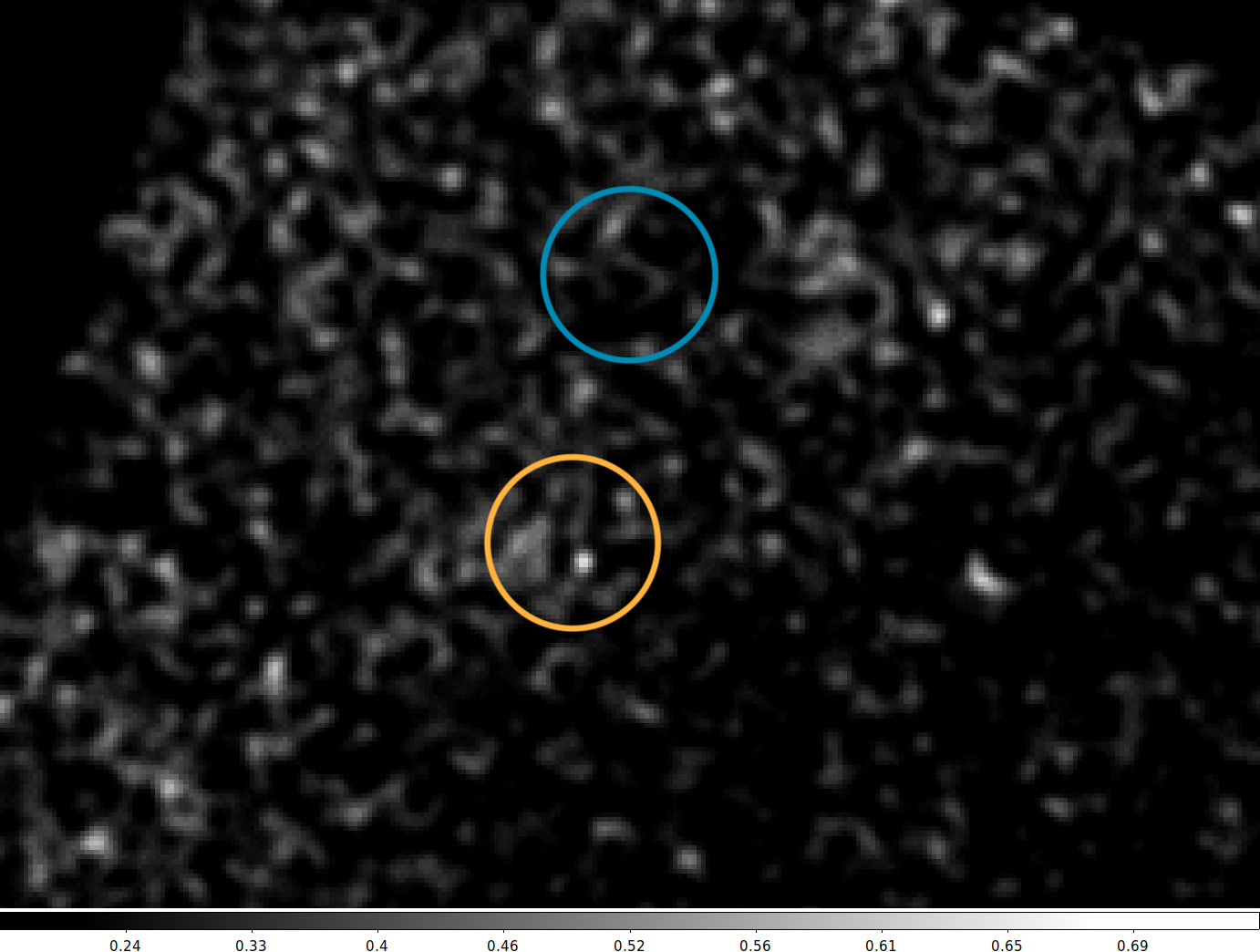}
\caption{Smoothed 3--24~keV NuSTAR FPMA image of the target J144013+024744. We define the source and background regions as circles with a 45.00$^{\prime\prime}$ radius. The source is represented by a solid orange circle while the source-free background is represented by a solid blue circle.}
    \label{fig:nustar}
\end{figure}

The target J1440 was observed on 2020-09-02 for $\approx 100$ ks with NuSTAR, with an obsid 60601028002. 
The data were processed using HEAsoft v6.29\footnote{https://heasarc.gsfc.nasa.gov/docs/software/heasoft/} while {\sc nupipeline} was used to produce cleaned and calibrated event lists. 
To account for the background enhancement due to the South Atlantic Anomaly, 
we set {\sc SAA$=$strict} and {\sc Tentacle$=$yes}. We selected the source and a source-free  background using 45.00$^{\prime\prime}$ radius circular regions centered at $(\alpha, \delta) = \rm(14h40m12.66s, +02d47m41.53s)$ and $\rm(14h40m09.47s, +02d49m49.56s)$, respectively (Figure ~\ref{fig:nustar}). {\sc Nuproduct} was used to extract the X-ray spectra of our target from the event files. We generated the spectra for the two Focal Plane Modules (FPM) onboard NuSTAR namely FPMA and FPMB. The net count rate in the 3-79~keV range for FPMA was $1.56 \times 10^{-3} \pm 4.76\times 10^{-4}$ $\rm counts~s^{-1}$, while that for FPMB was $1.63 \times 10^{-3} \pm 5.01\times 10^{-4}$ $\rm counts~s^{-1}$. The cleaned exposure times for FPMA and FPMB was 85.94 ks and 84.63 ks respectively. 
 
To measure the source significance, we followed the algorithm adopted by \citet{mullaney2015nustar,lansbury2017nustar}, which operates on the false-probability images generated based on the NuSTAR background and science images \citep[see Section 2.3 of][for further details]{mullaney2015nustar}.
The results for the source-detection algorithm for the 3--8~keV, 8--24~keV, 3--24~keV and 30--50~keV bands are summarized in Table~\ref{tab:detection}.

\begin{table}[t]
\footnotesize
\caption{\scriptsize NuSTAR source significance for J1440. To perform this algorithm we chose two annulus of sizes 90$^{\prime\prime}$ and 160$^{\prime\prime}$ for source and background respectively: 
(1) energy band in~keV. (2) FPM detectors on which the source detection algorithm was applied namely FPMA (A) and FPMB (B). 
(3) Src represents the total source + background counts.
(4) Bgd represents the total background counts in the background region. 
(5) $B_{src}$ represents the background counts estimate for the source region. 
(6) $P_{\rm False}$ is the binomial false probability that a signal is coming from random fluctuations rather than the source.}\label{tab:detection}
 \begin{tabular}{lcllll}
 \hline
 Band & FPM & Src & Bgd & Bgd$_{\rm src}$ & $\rm P_{\rm False}$ \\

(1) & (2) & (3) & (4) & (5) & (6) \\ 
 \hline
  3--8 & A+B & 292 & 712  &  196.6 & $1.3\times10^{-10}$ \\ 
 8--24 & A+B & 301 & 847 &   233.9 & $9.5\times10^{-5}$ \\
 30--50 & A+B & 233 & 820 &    226.5 &  $0.365$ \\
 3--24  & A+B & 592 & 1553  &  428.9 & $5.5\times10^{-14}$  \\
 3--24  & A & 261 & 594  &  164.1 & $1.9\times10^{-12}$ \\
 3--24  & B & 331 & 959 &   264.9 & $2.6\times10^{-4}$ \\
\hline
\end{tabular}
\end{table}

From  Table~\ref{tab:detection}, we can infer that the source has significant detections in the 3--8, 8--24, and 3--24~keV bands and is not detected at 30-50~keV. For the 3--24~keV band, where the source is detected most strongly, the source shows a stronger detection in FPMA over FPMB due to lower background.

\section{X-Ray spectral analyses}\label{sec:xray}

In this section, we analyze the NuSTAR spectra of J1440 by fitting them to phenomenological models of different complexity, including a simple absorbed power law and the Pexmon model \citep{pexmon} that also takes into account the Compton reflection and fluorescent iron emission lines. 
We also fit the data with a physical model RXTorus \citep[see \S~\ref{sec:rxtorus}][]{paltani2017reflex}. RXTorus model allows for variable metallicity, hence, can be useful in modelling X-ray spectrum of dwarf galaxies where the metallicity might be  significantly lower than more massive galaxies. This is also motivated by recent studies that suggested the metal abundance in the AGN torus is comparable to those in the $\sim 10^{2-3}$~pc scale narrow-line region \citep{hikitani2018}.
We note that the total X-ray counts for J1440 are limited. The inclusion of physical models in our analysis is to provide some insights into if the torus  models built for typical AGNs can potentially be extended to the lower $M_\bullet$ and $M_\star$ ends, as J1440 is among the few dwarf active galaxies with a nucleus in the IMBH class that are observable with current instruments.

\begin{figure*}
    \plottwo{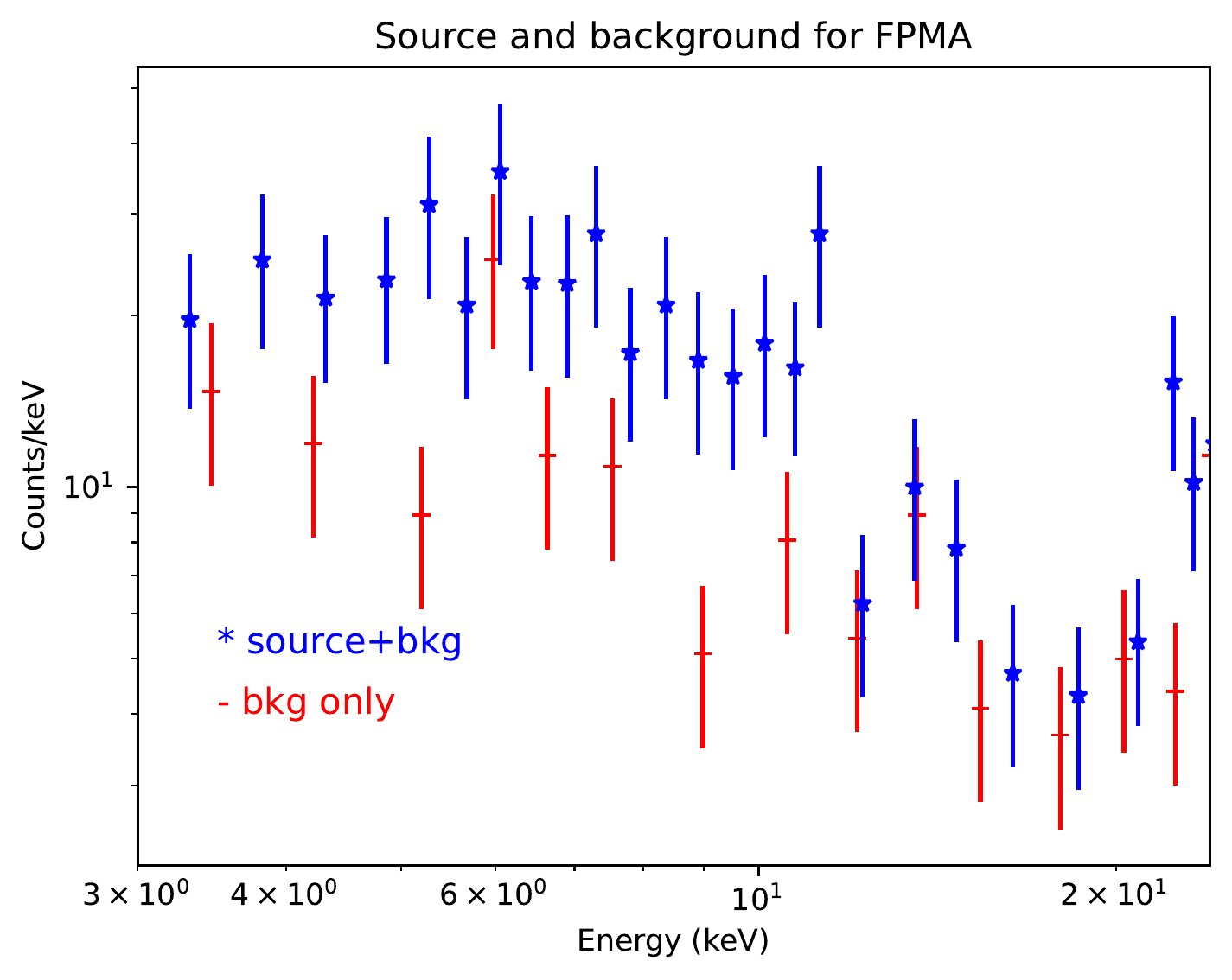}{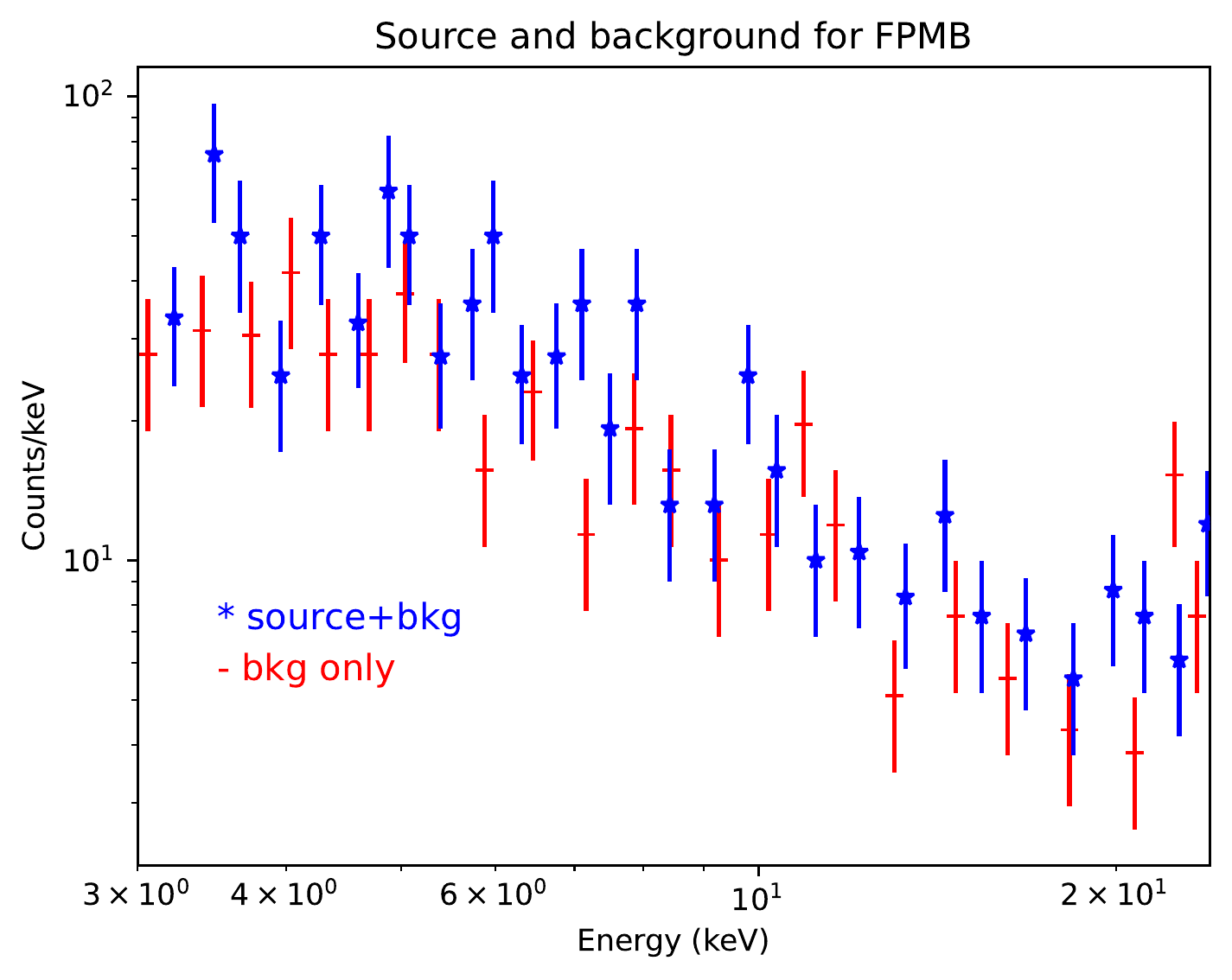}
    \caption{
    \footnotesize
    The source and background for the two FPM detectors. Source and background in both the plots are binned at 10 counts/bin for better display. The FPMB spectrum shows a significantly higher background counts at $<7~\rm keV$ than that of FPMA.} 
    \label{fig:databkg}
\end{figure*}

Since the data from FPMB is very heavily background dominated at $<7$~keV and  both the detectors show heavy background domination at $ > $ 24~keV (see Figure~\ref{fig:databkg}), we limit our analysis of NuSTAR data to only the FPMA detector in the 3--24~keV energy range. 
While there are archival soft X-ray data from Chandra \citep{baldassare2017x} and XMM-Newton \citep{thornton2009emission}, 
the signal-to-noise ratio for these data are too low at $\geq 1$~keV to include in our spectral analysis. 
We use the Bayesian X-ray analysis \citep[BXA,][]{2014A&A...564A.125B} software to fit our data. BXA utilizes a nested sampling algorithm\footnote{\url{https://johannesbuchner.github.io/UltraNest/}} to more effectively search for the likelihood maximum  when the model parameters exhibit degeneracy. We also utilized BXA's principle component analysis (PCA) based background model for the FPMA detector. The unbinned source spectrum and its background from the FPMA detector were simultaneously fitted (without subtracting) between 3--24~keV with different models using the Cash statistic. All these fits were performed with the  Sherpa fitting software \citep{2001SPIE.4477...76F}.
Regardless of our choice of fitting algorithm and software, the photon index in our fitting could not be constrained due to the lack of good-quality soft X-ray data, therefore, we fixed the photon index to $\Gamma =2.13$ (see Section 4.1 for justification of this choice based on multi-wavelength data).
In all the models, the metallicity was set to $\rm Z=0.006$ to account for the likely low metallicity typical for dwarf galaxies (note that the metallicity obtained from the SED fit in Sec.~\ref{sec:sed} was $\rm Z=0.004$, however, the minimum $Z$ allowed in RXTorus was $0.006$). For our analysis, we ran BXA with 400 live points with the fraction of the integration remainder set at 0.5. Other BXA parameters were set to their default values.

In this work we do not consider a joint analysis of NusTAR data and the archival XMM-Newton data.
This decision is primarily driven by the lack of $>1$ keV source photons in the XMM-Newton spectra, and the fact that the XMM-Newton spectrum was already well-fitted with a diffuse hot plasma model (see \S~\ref{sec:xmm}). Given the NuSTAR spectrum also has limited counts, a joint analysis that requires both the AGN component and a different soft X-ray component can actually introduce more parameters than the increased degree of freedom hence making the model parameters even less-constrained.

\subsection{Phenomenological models}
\begin{figure*}
     \centering
         \includegraphics[width=0.32\textwidth]{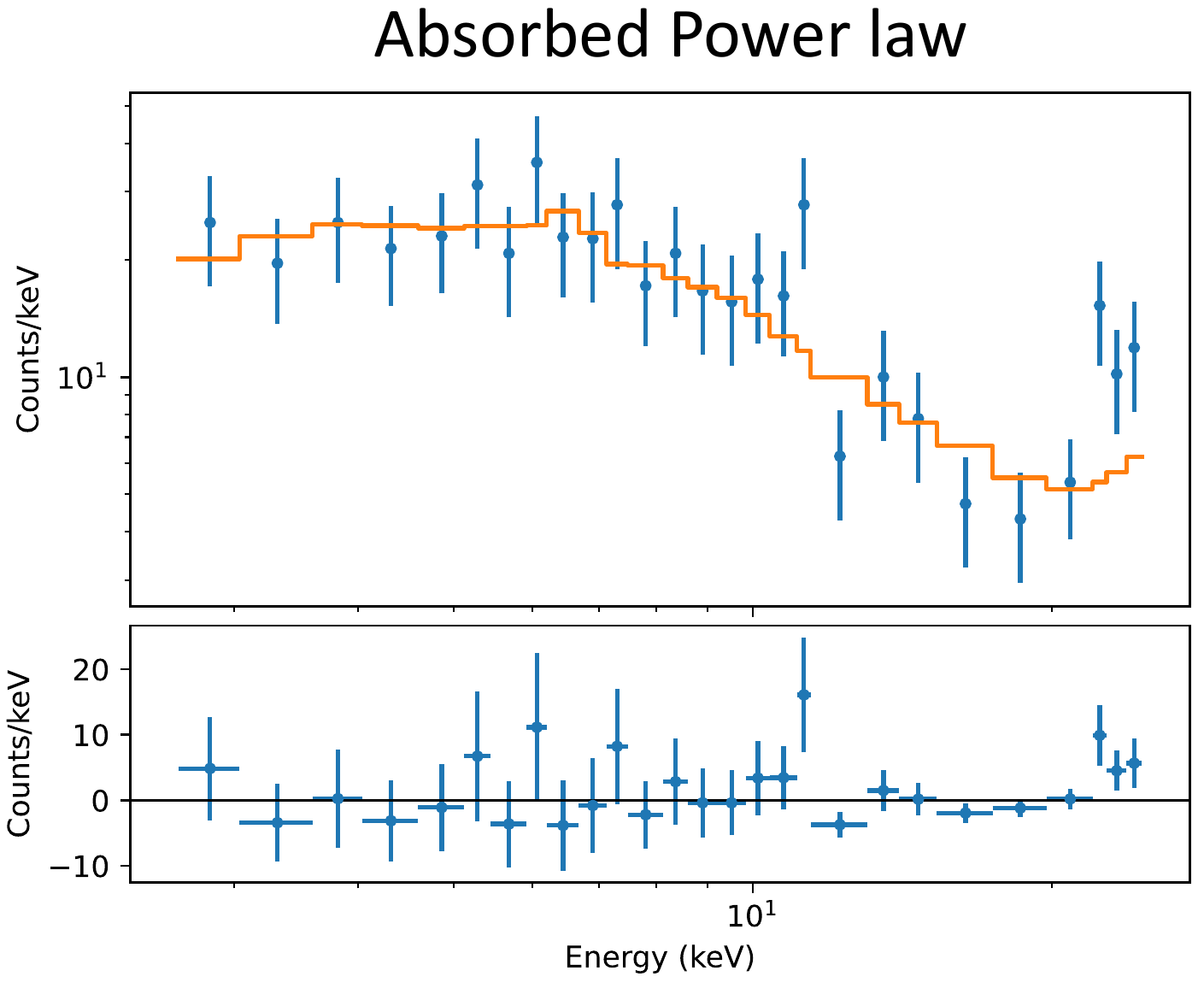}
         \includegraphics[width=0.32\textwidth]{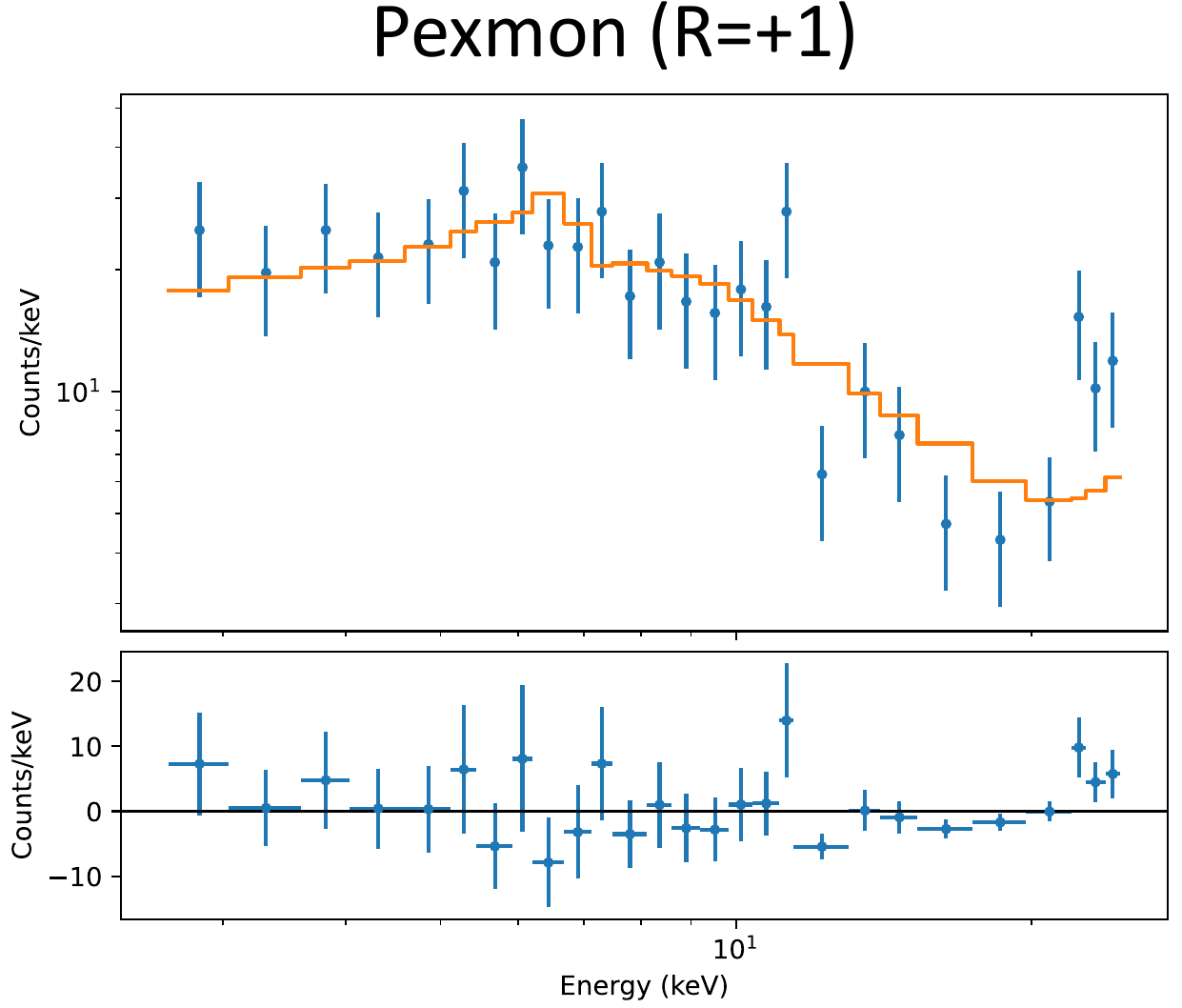}
          \includegraphics[width=0.32\textwidth]{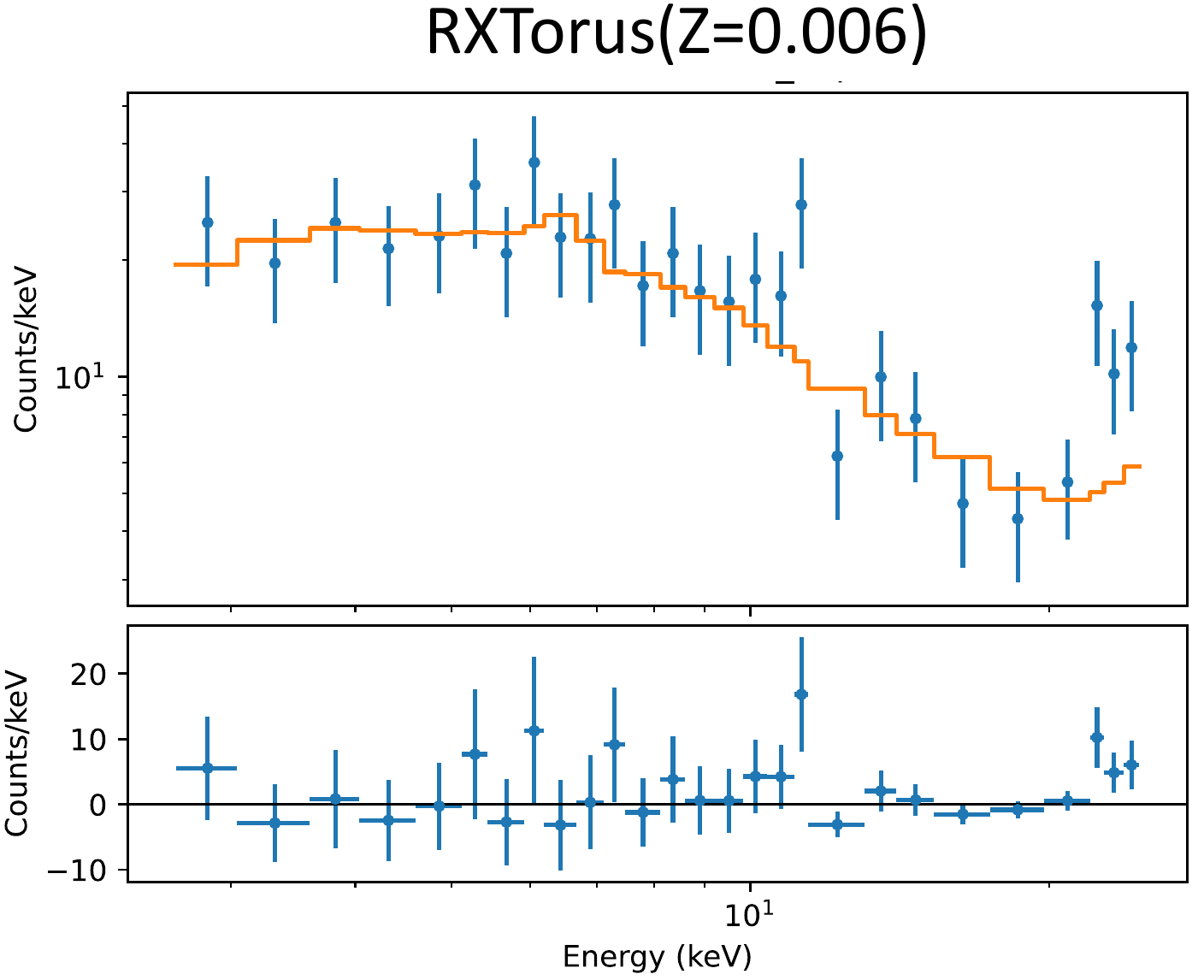}
        \caption{
        \footnotesize
        X-ray fitting spectrum for different models: (a) Absorbed Power law (b) Pexmon model with $\rm R=+1$ and (c) RXTorus model with $\rm Z=0.006$. The spectrum is binned at 10 counts per bin for better display.}
        \label{fig:abspex}

\end{figure*}

Before delving into complex physical models, we evaluate the parameters with a simple absorbed power-law model and Pexmon model.
The absorbed power law model in {\sc Sherpa} can be written as \texttt{xstbabs$\times$(xszphabs $\times$ xszpowerlw)}. Here, \texttt{xstbabs} accounts for Galactic absorption and was set to $2.9\times 10^{20} \rm cm^{-2} $.\footnote{The value of Galactic column density was calculated using the HEASoft w3nh tool at \url{https://heasarc.gsfc.nasa.gov/cgi-bin/Tools/w3nh/w3nh.pl}. We chose the column density obtained from the HI4PI map.} \texttt{xszphabs$\times$xszpow} is the redshifted absorbed power law.  
The best-fit power-law model parameters suggest the source to be heavily obscured with  $\log(\rm N_{\rm H}/\rm cm^{-2})= 23.54_{-0.28}^{+0.21}$  (0.16, 0.50 and 0.84 quantiles), and a fit statistic (Cstat/d.o.f.) of 507.35/521. The unabsorbed and the absorbed 2--10~keV X-ray luminosities for the absorbed power law model are $\log(L_{2-10} (\rm ergs~s^{-1}))= 41.41_{-0.15}^{+0.12}$ and $\log(L_{2-10} (\rm ergs~s^{-1}))= 40.93_{-0.10}^{+0.09}$ respectively (see Figure~\ref{fig:abspex} for spectral fit and residual). If metallicity (Z=0.006) is included in the absorbed power law model (using \texttt{xsvphabs}), the obscuring column density increases to $\log(\rm N_{\rm H}/\rm cm^{-2})= 24.04_{-0.28}^{+0.19}$.

\begin{figure}
    \includegraphics[width=\columnwidth]{ 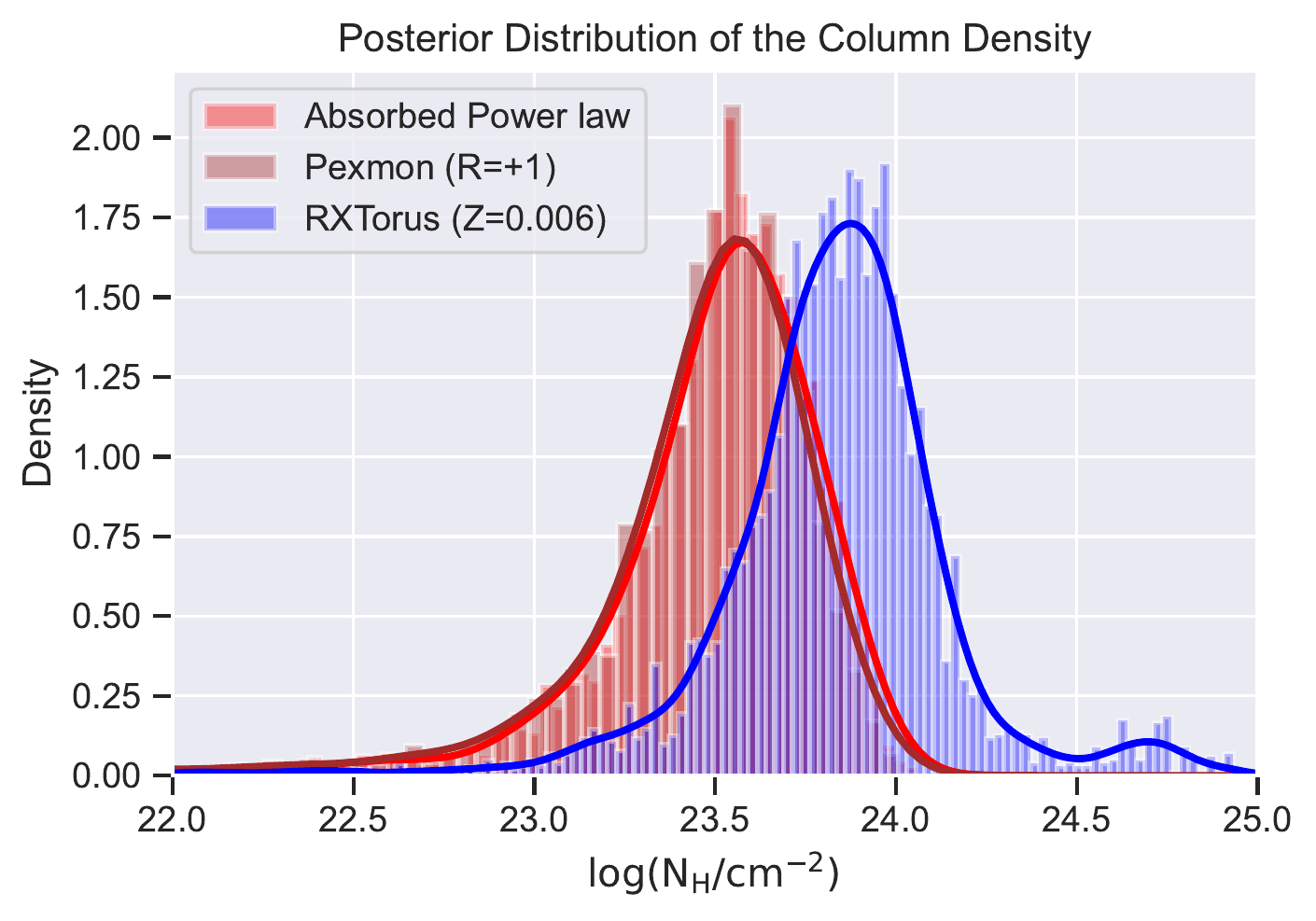}
    \caption{\footnotesize
        The posterior distribution of the column density for the three  models fitted to the NuSTAR data. RXTorus with $\rm Z=0.006$ produces the highest column density almost in the Compton-thick regime}
    \label{fig:posterior}
\end{figure}

We also modelled the X-ray spectrum using the Pexmon model which combines a power-law of a fixed cut-off energy with reflection from neutral Compton reflector and fluorescence lines. The model in {\sc Sherpa} can be written as \texttt{xstbabs$\times$(xszphabs $\times$ xspexmon)}. We set the inclination angle to $85^{\circ}$, though a different inclination angle did not significantly affect our results. Other parameters of \texttt{Pexmon} were set to their default values. The reflection coefficient of \texttt{Pexmon} was set to $\rm R=-1$ to simulate a reflection-dominated model while the cut-off energy was fixed at 400 keV
The column density of $\log(\rm N_{\rm H}/\rm cm^{-2})= 21.47_{-0.99}^{+1.16}$ was poorly constrained as we were unable to obtain a Gaussian posterior distribution. We then fixed the reflection coefficient to $R =+1$ to include the power law component. This configuration of Pexmon produced a well-constrained column density (see Figures \ref{fig:posterior}).
Since a simple absorbed power-law or a Pexmon model does not capture the physics of an obscuring torus, we use physical models to constrain the properties of a physical torus.

\subsection{RXTorus}\label{sec:rxtorus}
RXTorus is a model of obscuring torus which has a variable ratio of radius of minor to major axis of the torus. RXTorus model allows for different metallicities of the AGNs (ranging from $\rm Z=0.006$ to $\rm Z=0.04$), hence, can be used to model low metallicity AGNs particularly those in the dwarf galaxies. Moreover, it also allows for disentanglement of its line of sight column density and equatorial column density which it can measure up to $10^{25} \rm ~cm^{-2}$. The photon index can vary between $\Gamma=1$ and  $\Gamma=3$. The model in {\sc Sherpa} can be written as\footnote{https://www.astro.unige.ch/reflex/} \texttt{xstbabs $\times$ (RXTorus-cont-0.3 $\times$ xscutoffpl + RXTorus-rprc-0.3)}. Here, \texttt{RXTorus-cont-0.3 * xscutoffpl} is the absorbed power law component, \texttt{RXTorus-rprc-0.3} is the reprocessed component which includes scattering and fluorescence emission and 0.3 represents the metallicity in the unit of solar metallicity of these modules ($\rm Z=0.006$).\footnote{ The metallicity calculated from CIGALE and the Mass-metallicity relation was $\rm Z= 0.004$, however, the lowest metallicity allowed in RXTorus model is $\rm Z= 0.006$} In BXA, we created log-uniform priors for \texttt{RXTorus} normalization, column density and background normalization associated with the FPMA detector while a uniform prior was created for the inner to outer radius ratio of torus (r/R). The photon index, cut off energy, and column density of the all the modules were linked. The cutoff energy was set at 200~keV, the default value of the RXTorus model. 
The inclination angle was set to $90^{\circ}$ to simulate an edge-on viewing, however, we do not find significant changes in the posterior distribution for different inclination angles.

\begin{figure*}
    \includegraphics[width=0.48\textwidth]{ 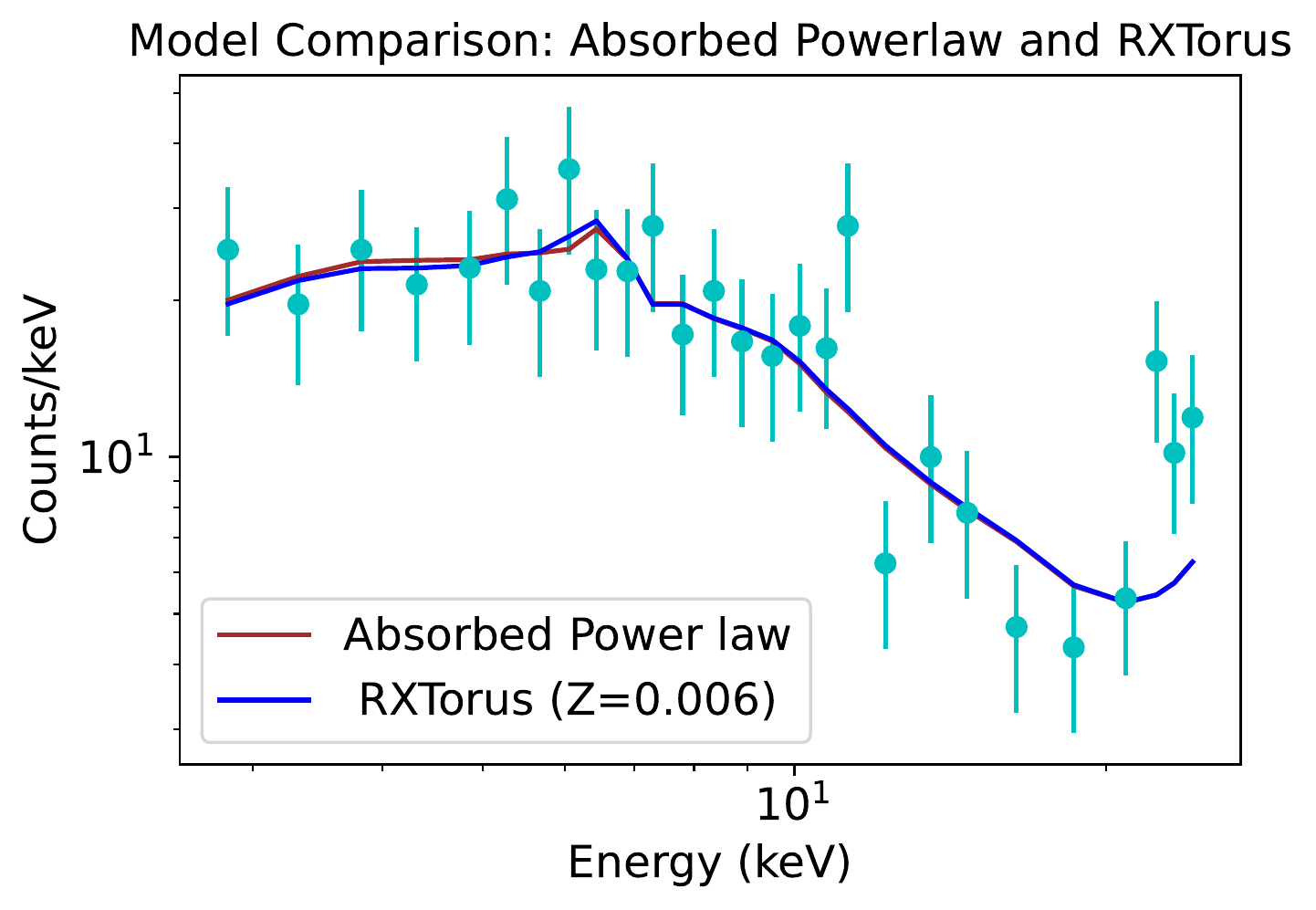} 
    \includegraphics[width=0.48\textwidth]{ 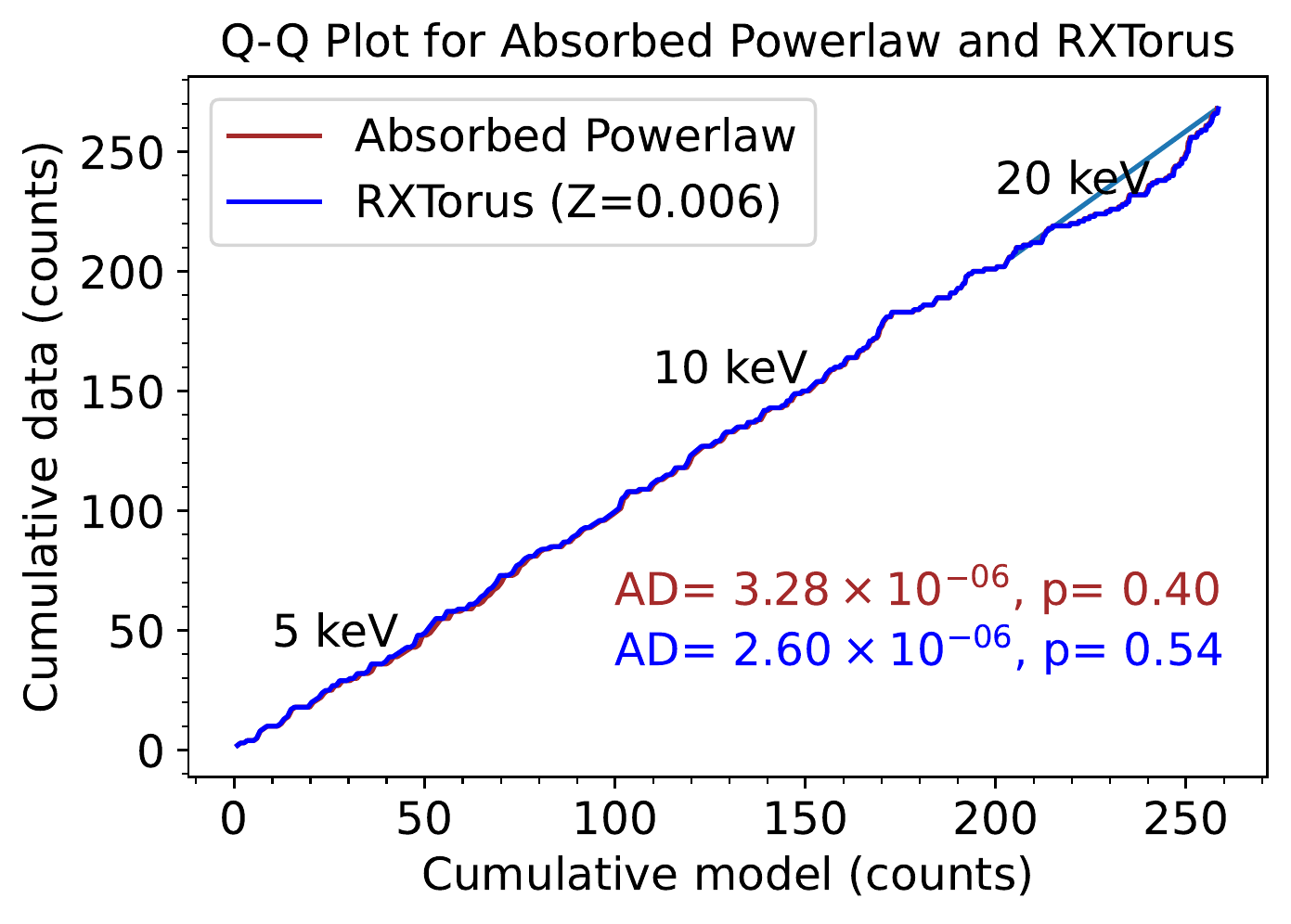}
    \caption{(left) Best-fit X-ray spectra for absorbed power-law and the physically motivated RXTorus model with $\rm Z=0.006$. Both these models explained the data reasonably well, however, a mismatch between data and model grows at higher energy ($>20$~keV). We binned the data to 10 counts/bin for better display. (right) The cumulative distribution of the unbinned data and the unbinned model is plotted in the Quantile-Quantile (Q-Q) plot from 3--24~keV. 
    The lower AD-stats and higher p-value for the RXTorus model with $\rm Z=0.006$ suggests this physical model might provide a slightly better fit statistics than the absorbed power-law only model. We note the difference is marginal and a higher quality spectrum would be required to confirm the presence of a reprocessed component from the putative torus.}
    \label{fig:QQ}
\end{figure*}

\subsection{X-ray fitting results}
We show the X-ray fitting results for the phenomenological and physical models in Figures \ref{fig:abspex}. The best-fit parameters of the models discussed in this work are listed in Table~\ref{tab:result}. We also show the posterior distribution of the best-fit column densities for all the models used in this work in Figure~\ref{fig:posterior}. All these models indicate the object to be heavily obscured. For the two phenomenological models simple absorbed power-law and Pexmon ($\rm R=+1$), the column densities are $\log(N_{\rm H}/\rm cm^{-2})$= $23.54_{-0.28}^{+0.21}$ and $23.52_{-0.29}^{+0.19}$, respectively. For the low-metallicity RXTorus model ($\rm Z=0.006$), the best-fit column density almost reaches the Compton-thick regime with $\log(N_{\rm H}/\rm cm^{-2})$= $23.85_{-0.25}^{+0.22})$. The log-likelihood function which BXA uses to compare models were consistent within margins of error.
 We do not find significant changes in the column density when a variable inclination angle is used. However, a variable photon index leads to a poor constraint on the column density; hence it was fixed to $\Gamma=2.13$.
We note the default cut-off energies are not the same between the
physical model and the phenomenological models. However, 
as demonstrated by \cite{balokovic2020nustar}, the intrinsic cut-off energies span a wide range from 140--500~keV for 68\% of their large sample of AGNs with high-quality NuSTAR and Swift/XRT data. Also, the spectral fit statistics for their sample do not improve significantly even if the cut-off energy was allowed to vary freely. Therefore, the different cut-off energies are unlikely to qualitatively change our results.
 
\begin{figure*}
    \plottwo{ 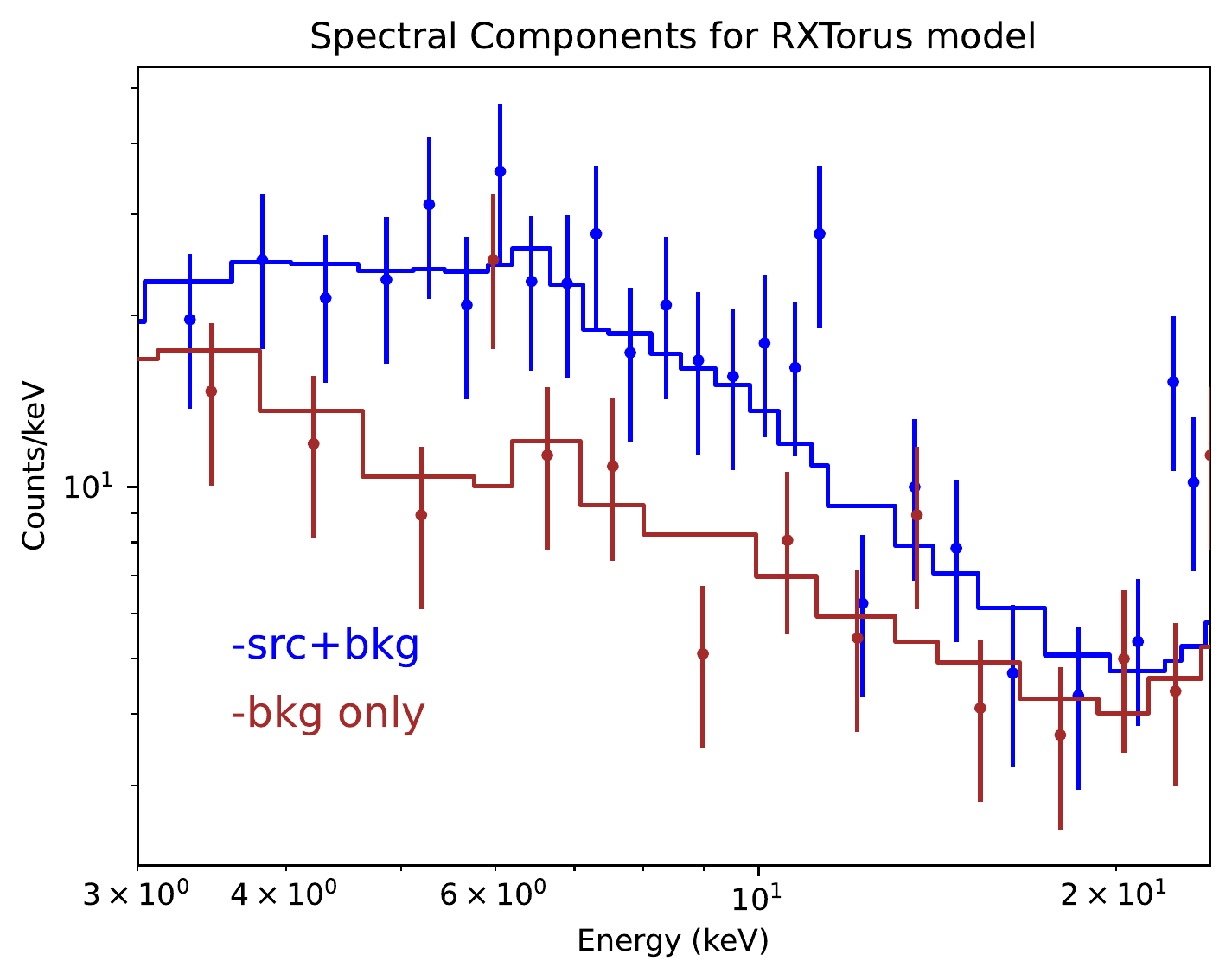}{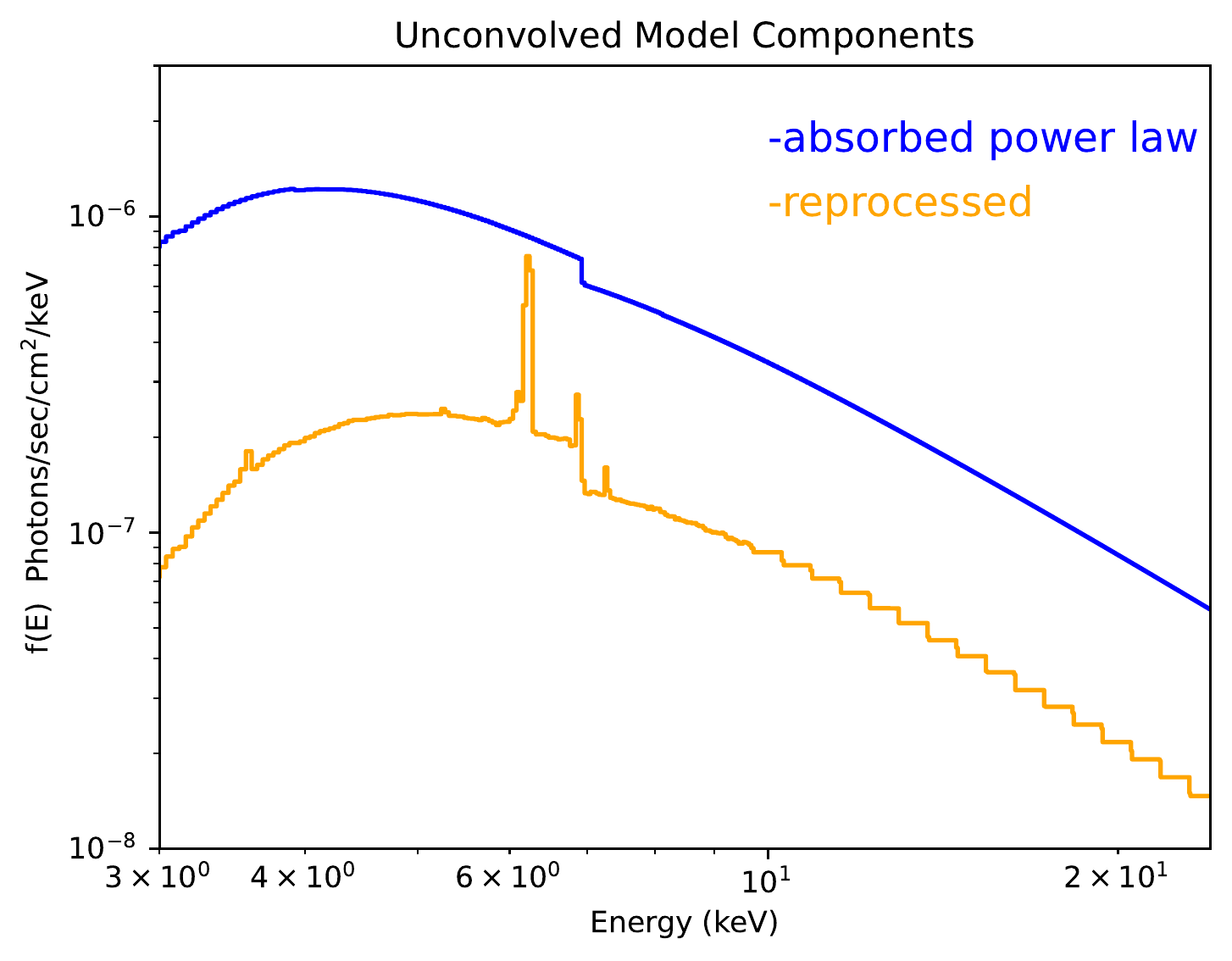} 
    \caption{\scriptsize \textbf{Spectral Components} (left) Result of simultaneous spectral fitting for the RXTorus model depicting the source $+$ background model and the background model only. (right) To assess the contribution of each of the additive model components, we plotted the unconvolved source spectrum (without the background model) for the RXTorus model. Although the source model is dominated by absorbed power law component, we see non-trivial contribution from the scattered component particularly in 6.4~keV region. However, the current data prevents us from constraining the width of this Fe line. }
    \label{fig:component}
\end{figure*}

We also show the combined model plot (binned for better display) and the Quantile - Quantile plot (Q-Q plot) of the unbinned data to compare between the absorbed power-law and the RXTorus model (Figure~\ref{fig:QQ}). All these models explain the data acceptably well, but the models deviate from the data at higher ($\gtrsim 20$~keV) energies, possibly due to the high background noise. To quantitatively assess the goodness of the fit, we performed a two sample Anderson-Darling (AD) test \citep{anderson1954test}. A two sample Anderson-Darling test measures the distance between the two cumulative distribution by estimating the square of the difference of the two functions multiplied with a weight function. Since the model was derived from the data, we performed bootstrapping with 2000 iterations to estimate the p-value. These values are quoted in Figure~\ref{fig:QQ}--right. The lower AD stat and a higher p-value indicates that the data and model are drawn from the same distribution and corroborates our findings from the Q-Q plot that RXTorus with its sub-solar metallicity provides the best fit to the data. In Figure ~\ref{fig:component} (left) we show the spectral fitting results for the source+background and background only models. Both the source and background model seems to explain the data reasonably well.
The unconvolved model components (Figure ~\ref{fig:component} (right)) for the RXTorus models show that the overall spectrum is dominated by an absorbed power-law component; however, a non-negligible scattered and fluorescence component is present in the best-fit spectra of physical models. We reiterate that we did not make use of the soft X-ray data from XMM-Newton nor Chandra, and did not include an additional component to account for the $<3$~keV emission due to the lack of high-quality soft X-ray data, and the fact that the $>3$~keV emission for this source cannot be explained with typical diffuse X-ray emission components such as {\sc apec}. 
A high-quality $<3$~keV X-ray spectrum is needed to further constrain the spectral properties such as the intrinsic photon-index, and the strength of the scattered and reflected components. 

To assess if the X-ray luminosity of J1440 might have a non-AGN origin, we calculate the expected luminosity from high-mass (HMXB) and low mass X-ray (LMXB) binaries. We used the empirical relation derived by \citet{lehmer2010chandra} along with the SED-derived SFR and stellar mass discussed in Section~\ref{sec:sed} and find $L_{\rm LMXB }=(1.4\pm 0.2)\times10^{38}$ $\rm ergs~s^{-1} $ and $L_{\rm HMXB }=(6.8\pm4) $ $\rm ergs~s^{-1} $. 
We compare this with the 2--10~keV intrinsic luminosity for each model (see Table~\ref{tab:result}) calculated using {\sc Sherpa}'s $``\rm calc$\textunderscore energy\textunderscore $\rm flux"$ command. This command was applied to the  power-law component of the model to obtain the intrinsic luminosity. We find that the intrinsic 2--10~keV X-ray luminosities (Table~\ref{tab:result}) are $\gtrsim 2$ orders of magnitude higher than that expected from X-ray binaries (XRB).
Therefore, the X-ray spectrum of J1440 is unlikely to suffer from significant XRB contamination. 

We summarize the key results of our X-ray spectral analysis below:

\begin{enumerate}
    \item The intrinsic X-ray luminosities for the models discussed in this work are much higher than the expected contribution from X-ray binaries, which strongly supports the presence of a hard X-ray AGN despite the faint soft X-ray fluxes found in the literature \citep[][]{2009ApJ...705.1196T,baldassare2017x}.
    
    \item All the models indicate the object to be heavily obscured.
    For the simple absorbed power-law and Pexmon (R$=$+1) model, the column densities are $\log(N_{\rm H}/\rm cm^{-2})$= $23.54_{-0.28}^{+0.21}$ and $23.52_{-0.29}^{+0.19}$, respectively
    For the RXTorus model with a low metallicity, the best-fit column density nearly reaches the Compton-thick regime with $\log(N_{\rm H}/\rm cm^{-2})$= $23.85_{-0.25}^{+0.22}$.
    \item The geometry of the torus was poorly constrained as we were unable to obtain a Gaussian posterior distribution for the geometry parameter in the RXTorus model. The 0.16, 0.50 and 0.84 quantiles for the ratio of inner to outer radius parameter for the RXTorus model was $0.47_{-0.25}^{+0.29} $. We need more good quality data to constrain the geometry of the obscuring torus.
    
    \item Qualitatively, the spectra of J1440 is dominated by the absorbed power-law component based on the results from the physical model, but high-quality soft X-ray data and higher signal-to-noise ratio NuSTAR spectra are needed to further constrain the spectral properties. 

\end{enumerate}

\begin{table*}
\footnotesize
\centering

\caption{\textbf{ Posterior estimates for different models}} The range corresponds to 0.5 quantile with upper bounds corresponding to 0.84 quantile while lower bound corresponds to 0.16 quantile. Here, $\log{\rm (Norm_{zero})}$ represents the normalization for the zeroth order power-law,  $\log(\rm Norm_{PCA})$ is the normalization for the PCA based background model associated with FPMA detector.
\label{tab:result}
\newline

 \begin{tabular}{cccc} 
 \hline
  Model & Abs.Powerlw &Pexmon($\rm R=+1$) & {RXTorus} \\
  \hline\hline
  $\log(\frac{\rm N_{\rm H}}{\rm cm^{-2}})$ & $23.54_{-0.28}^{+0.21}$  & $23.52_{-0.29}^{+0.19}$ & $23.85_{-0.25}^{+0.22}$  \\ 
  \hline
  $\log{(\rm Norm_{zero})}$ & $-4.15_{-0.15}^{+0.12}$ & $-4.22_{-0.14}^{+0.12}$ & $-3.98_{-0.20}^{+0.25}$  \\
 \hline
 $\log(\rm Norm_{PCA})$  & $2.74_{-0.03}^{+0.03}$& $2.74_{-0.03}^{0.03}$ & $2.74_{-0.03}^{+0.03}$        \\ 
 \hline
$\rm (CStat/dof)_{\rm source+bkg}$ & $508.39/521$ & $510.14/526$  & $510.56/521$ \\
\hline
$(\rm CStat/dof)_{\rm bkg~only}$ & $393.91/521$  & $393.36/521$  & $393.34/521$ \\
\hline
$\log(\rm L_{2-10}$ ($\rm ergs~s^{-1} $)) & $41.41_{-0.15}^{+0.12}$ & $41.38_{-0.14}^{+0.12}$  & $41.62_{-0.19}^{+0.23}$ \\
\hline
$\log(Z)$ & $-461.65\pm 0.47$ & $-461.12\pm0.44$  & $-460.26\pm0.46$ \\
\hline
\end{tabular}
\end{table*}

\section{Multi-wavelength properties of J1440}

In this section we supplement our findings from the X-ray spectral fitting with the multi-wavelength observations of J1440.

\subsection{Constraining photon index from the Eddington ratio}
As discussed in Section~\ref{sec:xray}, our data quality prohibits us from exploring the full range of the X-ray power-law photon index $\Gamma$. We choose the value informed by the multi-wavelength data, specifically, the AGN bolometric luminosity. 
For an accreting SMBH, its bolometric luminosity is tied to the global accretion-rate that can affect the seed photon emission and can modulate the X-ray spectral slope \citep[e.g.,][]{shemmer2006hard,ishibashi2010x}. A linear relation between Eddington ratio ($\lambda$) and photon index has been documented in the past for systems accreting at high Eddington ratios \citep[e.g.,][]{yang2015correlation, brightman2013statistical, Trakhtenbrot2017BATAS, shemmer2008hard,shemmer2006hard}. In \cite{brightman2013statistical}, the relation can be expressed as $\Gamma = (0.32 \pm 0.05)\log \lambda+(2.27 \pm 0.06)$. 
For J1440, the bolometric luminosity derived from $[O \rom {3}]$(5007\text{\AA}) is 
$\log(L_{\rm bol}/{\rm erg\ s}^{-1})= 43.96 \pm 0.60$ based on the \citet{sternlaor2012} empirical relation.
Similarly, the  bolometric luminosity derived from $[O \rom {4}](25.89\rm \mu m$) based on \citet{goulding2010towards} is $\log(L_{\rm bol}/{\rm ergs\ s}^{-1})= 42.94 \pm 0.40$. We also calculated the bolometric luminosity directly by integrating the best-fit SED described in Section~\ref{sec:sed} to be $\log(L_{\rm bol}^{\rm SED}/{\rm erg\ s}^{-1})= 43.64\pm 0.04$. These values are consistent with each other within the $\sim 1\sigma$ uncertainty.
Using the bolometric luminosity from $[O \rom {4}]$(25.89$\rm \mu m$) and the black-hole mass obtained from broad $\rm H_{\alpha}$ line, we estimate the Eddington ratio of J1440 to be $\lambda$ $\sim$ 0.37. This translates to a photon index of $\Gamma = 2.13 \pm 0.06$ under the assumption of the $\Gamma-\lambda$ relation from \citet{brightman2013statistical}.  

A higher photon index for high-Eddington ratio, lower $M_\bullet$ systems has been found in the literature \citep[e.g.,][]{porquet2004xmm,done2012intrinsic,baldassare2017x}, as low $M_\bullet$ systems may have a higher accretion disk temperature with thermal emission extending into the soft X-ray energies. Moreover, one can argue that the  X-ray photon index could be softer as J1440 shows the presence of ${[ O \rom {4}](25.89\rm \mu m)}$ emission-line which requires an abundance of EUV photons to ionize and the presence of these EUV photons have been linked to soft X-ray fluxes (\citealt{timlin2021controls}; \citealt{telfer2002rest}).

\subsection{Comparison between different AGN luminosity indicators}
\begin{figure}
    \centering
    \includegraphics[width=0.5\textwidth]{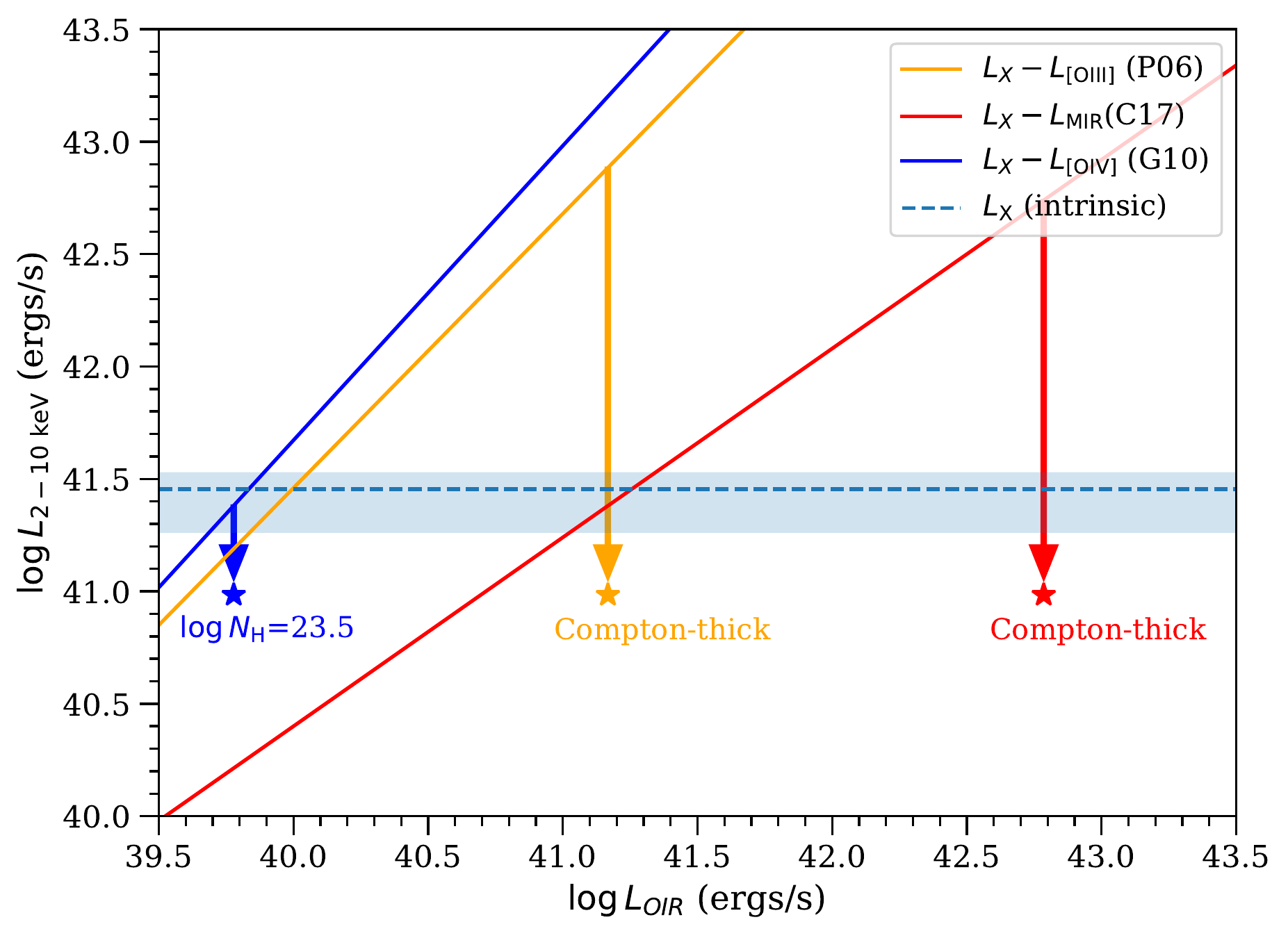}
    \caption{
    A comparison between the scaling relations between the intrinsic 2--10~keV luminosity and various AGN luminosity indicators derived based on samples of unobscured AGNs, including $L_{\rm 2-10 keV}$ vs. $L_{[ O \rom {3} ]}$ \citep[orange,][P06]{panessa2006x}, $L_{\rm 2-10 keV}$ vs. $L_{6\rm \mu m}$ \citep[red,][C17]{chen2017x}, and $L_{\rm 2-10 keV}$ vs. $L_{[ O \rom {4} ](25.89\rm \mu m)}$ \citep[blue,][G10]{goulding2010towards}. 
    The observed $L_{\rm 2-10 keV}$ and the relevant optical-IR luminosities ($L_{\rm OIR}$) are shown as the stars. 
    The intrinsic X-ray luminosity (based on a simple absorbed power-law with metallicity set at solar metallicity) derived using the NuSTAR observations are shown as the horizontal dashed line with 1$\sigma$ uncertainties marked as the shaded region. For large surveys, a common practice is to use the deviation between the observed X-ray luminosity and intrinsic X-ray luminosity based on the scaling relations to estimate the obscuring column density. For J1440, the large deviations between the observed $L_{\rm 2-10 keV}$ and $L_{[ O \rom {3} ]}$ or $L_{6\rm \mu m}$ imply the target to be obscured by Compton-thick materials, while the $N_{\rm H}$ value inferred from the $L_{\rm 2-10 keV}$ vs. $L_{[ O \rom {4} ](25.89\rm \mu m)}$ relation is consistent with the X-ray fitting results. 
    }
    \label{fig:lumin}
\end{figure}

While X-ray spectral analysis is an ideal way to confirm the presence of obscuring material in an AGN, 
the scarcity of sensitive X-ray instruments has limited such observations to only a small number of sources with deep X-ray observations.
One commonly adopted practice to find candidate X-ray obscured AGN in large surveys is to compare the observed X-ray luminosity and AGN luminosity indicators at other wavelengths, assuming these different AGN luminosity indicators follow simple scaling relations derived for unobscured type 1 sources. 
In Figure~\ref{fig:lumin}, we compared the 2--10~keV intrinsic luminosity  with that from the ${[ O \rom {3} ](5007)\text{\AA}}$ \citep{barth2008low}, the ${[ O \rom {4}](25.89\rm \mu m)}$ (H17) and the 6$\rm \mu$m (estimated from the SED fits) mid-IR luminosity. The solid orange line represents the $L_{2-10}~\rm keV$ vs $L_{[ O \rom {3} ](5007)\text{\AA}}$ relation obtained by \cite{panessa2006x}, where the linear relation can be expressed as $\log (\frac{L_{2-10~\rm keV}}{\rm ergs~s^{-1}})= (1.22 \pm 0.06) \log(\frac{L_{[O\rom{3}(5007\text{\AA})}}{\rm ergs~s^{-1}} + (-7.34 \pm 2.53)$. 
Similarly the solid red line shows the  $L_{2-10}$ vs. $L_{6\rm \mu m}$ relation obtained by \citet{chen2017x}. 
This equation can be expressed as $\log (\frac{L_{2-10~\rm keV}}{\rm ergs~s^{-1}}) = (0.84 \pm 0.03) \times \log (\frac{L_{6\rm \mu m}}{10^{45} \rm ergs~s^{-1}}) + (44.60 \pm 0.01)$. The solid blue line represents the $L_{\rm bol}$ vs. $L_{[ O \rom {4} ](25.89\rm \mu m)}$ relation obtained by \citet{goulding2010towards}, which can be expressed as $\log(\frac{L_{\rm bol}}{10^{44}\rm ergs/s})=(0.38 \pm 0.09) +(1.31 \pm 0.09)\times log( \frac{L_{[ O \rom {4}](25.89\rm \mu m)}}{10^{41}\rm ergs~s^{-1}})$. 
Here we use a X-ray bolometric correction factor of 25 \citep{brightman2017x} to convert the relation between $L_{\rm 2-10 keV}$ and $L_{[ O \rom {4}]}$.

With the intrinsic X-ray luminosity and $N_{\rm H}$ measured for J1440 with NuSTAR, we can assess how effective these different AGN luminosity indicators can be when used to identify heavily obscured AGNs when combined with X-ray observations. For this purpose, we calculated the absorbed 2--10~keV X-ray luminosity by fitting the FPMA data with a simple absorbed power-law model mentioned in Section 3.1, which leads to $\log(\frac{ L_{\rm 2-10 keV}^{\rm abs}}{\rm ergs~s^{-1}})=40.93$. 
If we assumed the intrinsic AGN X-ray luminosity follows the aforementioned empirical relations, 
$N_{\rm H}$ values estimated using the ratio between $L_{\rm 2-10 keV}^{\rm abs}$ and $L_{[O\rom{3}]}$ or $L_{6\rm \mu m}$ would be at least an order of magnitude higher than what we obtained from X-ray spectral fitting analysis. On the other hand, the ratio between $L_{\rm 2--10~keV}^{\rm abs}$ and ${[O\rom{4}]}(25.89\rm \mu m)$ predicts a column density of $ \log(\frac {\rm N_{\rm H}}{\rm cm^{-2}}) = 23.4$ ,  which is more consistent with the best-fit $N_{\rm H}$ with phenomenological and physical models described in Section~\ref{sec:xray}. The results are shown in Figure~\ref{fig:lumin}. 
One plausible explanation is that the higher ionization potential of ${[O\rom{4} ]}$ (59.4 eV) makes it a better AGN luminosity tracer in dwarf galaxies due to the more significant host-galaxy contamination effects in lines with lower ionization energy such as ${[O\rom{3} ]}$ (35 eV) \citep[e.g.,][]{melendez2008new}. 
As for $L_{6 \rm \mu m}$, we see non-negligible contributions from dust emission associated with stellar activity in the mid-IR wavelength of the best-fit SED. Given the limited photometric coverage in the mid-IR, it is possible that the best-fit AGN SED is still contaminated by the host-galaxy contribution. 
Moreover, if ${[O\rom{4}]}(25.89 \rm \mu m)$ lines are thought to be the better measure of the bolometric luminosity, then CIGALE seems to be over-predicting the bolometric and consequently $6 \rm \mu m$ luminosities. 

\subsection{Optical to X-ray luminosity ratio}
The ratio between the optical-UV to X-ray luminosity for AGNs has been actively studied in the past, particular in the parameterized form between the luminosity densities at $2500$\text{\AA} and 2~keV: $\alpha_{\rm OX}= -0.383 \log(\frac{L_{2500\rm\text{\AA}}}{L_{\rm 2keV}})$ \citep{tananbaum1979}. For typical, type 1 AGNs, a clear correlation has been established between $\alpha_{\rm OX}$ and $L_{2500\rm \text{\AA}}$ \citep[e.g.,][]{just2007x,lusso2010}, suggesting typical AGNs share a common radiation mechanism. However, it is not clear if this relation extends to AGNs powered by less-massive black holes in dwarf galaxies. For instance, the dwarf galaxies in the \citet{baldassare2017x} sample generally deviate from the linear relation, which was primarily attributed to the difficulties in separating AGN emission from that of the host-galaxy in such systems \citep[see Sec. 3.2 of ][]{baldassare2017x}. The target studied in this work, J1440, is a part of the \citet{baldassare2017x} sample and was found to have an $\alpha_{\rm OX}$ lower than the empirical $\alpha_{\rm OX}-L_{2500\text{\AA}}$ relation for typical AGNs by at least $\approx 0.6$. Here we recalculate $\alpha_{\rm OX}$ for J1440 using the intrinsic 2~keV luminosity based on the best-fit model using NuSTAR data. The UV-Optical emission of J1440 is heavily suppressed as shown in Figure~\ref{fig:SED}, therefore, we calculated $L_{2500\text{\AA}}$ using the power law relation $f_{\nu}\propto\nu^{-0.44}$, where we utilized the flux at $5100$\text{\AA} estimated from the $\rm H_{\alpha}$ emission-line (see section 4 of \citealt{dong2012x} for details of this method). 

We find an updated $\alpha_{\rm OX}$ value for J1440 of $\alpha_{\rm OX}=-1.26$. This is consistent with the value expected from the \citet{just2007x} relation within the margin of error. 
We attribute the different $\alpha_{\rm OX}$ values between our work and that from \cite{baldassare2017x}, $\alpha_{\rm OX}=-1.6$, to the presence of AGN obscuration for this object, and the soft X-ray data observed on Chandra and XMM-Newton might have been due to stellar activity of the host galaxy or photo-ionization of the gas in the AGN narrow-line regions, hence,  only the hard X-ray data can give a reliable estimated of intrinsic 2--10 KeV X-ray luminosity for heavily obscured AGNs due to its higher penetrating power. At least for J1440, the IMBH-powered AGN appears to have an optical-UV to X-ray spectral slope similar to typical AGNs. We note the $L_{2500\text{\AA}}$ we estimated using empirical relations likely has large uncertainties, but it is consistent with $L_{2500\text{\AA}}$ value measured using HST photometry by \cite{baldassare2017x}, albeit the direct photometry measurement is also likely highly uncertain due to host galaxy contamination.
Previous studies of low-mass AGN samples such as \citet{dong2012x} and \citet{baldassare2017x} have found a significant fraction of their sample to deviate away from the empirical $\alpha_{\rm OX}-L_{2500\text{\AA}}$ relation. The results from J1440 showed here suggest at least some of the low-mass AGN still follow the typical $\alpha_{\rm OX}$ relation when 
obscuration and host-galaxy contamination are properly corrected. A NuSTAR follow up of these low mass AGNs is thus warranted to study the extension of $\alpha_{ox}$ in low mass regimes.

\subsection{Connecting the IRS spectrum with the X-ray data}
Recently \citet{fernandez2021x} studied the IR emission-line ratios for AGNs selected from the Spitzer IRS archive. 
They found that the ratio between emission-lines of different ionizing energies can be used as a proxy of the ``hardness'' of the ionizing source (i.e., $[Ne\rom{2}](12.81 \rm \mu m$)/($[Ne\rom{2}](12.81 \rm \mu m$)+$[O\rom{4}](25.89\rm \mu m)$), see their $\rm Eq.1$). The EUV photons from the accretion disk in rapidly accreting systems can excite high-excitation lines such as $[O\rom{4}](25.89\rm \mu m)$ from the accretion disk, while the low-excitation lines such as $[Ne\rom{2}](12.81 \rm \mu m$) become more significant in systems that lack the strong UV continuum as they are accreting in a lower state. 
This transition between different accretion states are commonly seen in XRBs as their X-ray spectral hardness and count rates cycle through different accretion states \citep[e.g,][]{2004MNRAS.355.1105F}.
However, these transitions are not well studied in the case of AGNs, though, some studies suggests that changing look quasars might be a manifestation of change of state similar to XRBs \citep{noda2018explaining}. 
Despite potential uncertainties associated with original sample selection bias as well as local physical conditions, the mid-IR line ratio provides a useful tool for inferring the ``hardness'' for AGNs with heavily suppressed and/or soft X-ray emission dominated by host galaxy stellar activity.
With the Spitzer IRS emission-line measurements from H17\footnote{For J1440 $[O\rom{4}](25.89\rm \mu m)$ and $[Ne\rom{2}](12.81 \rm \mu m$) fluxes are $2.77\pm0.65$  and $3.64 \pm 0.70$, respectively.}, we can calculate the the line ratio defined above to be $\approx 0.6$. Since the Eddington ratio of J1440 is high ($\lambda_{\rm Edd}\approx 0.37$), it could imply that J1440 is currently in the soft accretion state and is transitioning into the hard accretion state. It is possible that $[Ne\rom{2}](12.81 \rm \mu m$) has more contamination from the host-galaxy for a dwarf galaxy such as J1440.
Future spatially resolved observations with JWST's MIRI might help with determining the nuclear line and continuum fluxes which will reveal more information regarding its accretion mechanisms.

\section{Conclusion and discussion}
In this paper we studied J$144013+024744$, which is an ideal candidate for an obscured AGN in a dwarf galaxy due to the  presence of the high-excitation $[O \rom {4}]$(25.89$\rm \mu m$) emission-line, an AGN marker that is less affected by host-galaxy contamination. 
We find hard X-ray emission in the NuSTAR data of J1440 consistent with that from an AGN, 
despite the faint soft X-ray emissions from the archival XMM-Newton ($\log(L_{2-10}/\rm ergs~s^{-1})=40.53\pm0.11$) and Chandra data ($\log(L_{2-10}/\rm ergs~s^{-1})= 38.79-40.36 ~(90\% ~\rm confidence)$).
We fitted the NuSTAR data from the FPMA detector with various physical and phenomenological models
and found the object J1440 to be a heavily obscured AGN. 
The main results are summarized below.
\begin{itemize}
  \item The SED fitting with UV-optical-IR photometry confirms the stellar mass of J1440 ($M_{\star}= 10^{9.20 \pm 0.05}M_{\odot}$) is indeed in the dwarf galaxy class, 
  and the fitting result suggests that an AGN component with a bolometric luminosity of $\log(L_{\rm bol}^{\rm SED}/{\rm erg\ s}^{-1})= 43.64\pm 0.04$ is needed to explain the broad-band SED of J1440.
  \item The bolometric luminosity estimated from $[O\rom {4}]$(25.89$\rm \mu m$) indicates that J1440 is accreting at a relatively high rate as its bolometric luminosity was close to 37$\%$ of its Eddington luminosity. The high Eddington ratio makes spectral analysis possible with NuSTAR observations despite the low BH mass of the AGN in J1440.
  \item We fitted the NuSTAR data with phenomenological and physical models. All these models provided a good fit of the NuSTAR data as determined from the quantitative Anderson-Darling test
  as well as qualitative tests such as the Quantile-Quantile plots shown in Fig.~\ref{fig:QQ}. The effect of metallicity might be important for low-metallicity systems such as dwarf galaxies as we noticed a significant increase in the absorbed column density with the RXTorus model.
  \item We compared the intrinsic 2--10 keV X-ray luminosity derived using NuSTAR observations ($L_{2-10}^{\rm int} \sim 10^{41.41}$~erg~s$^{-1}$) with AGN luminosities at other bands, including 
  $L_{[O\rom {3}](5007\text{\AA})}$, $L_{[ O \rom {4}](25.89\rm \mu m)}$ and $L_{6\rm \mu m}$. 
  We find that only $L_{[ O\rom{4}](25.89\rm \mu m)}$ traces the intrinsic X-ray luminosity of J1440 in a similar fashion as the empirical 
  $L_{[ O\rom{4}](25.89\rm \mu m)}-L_{\rm X}$ relation for unabsorbed AGNs. 
  This suggests the possibility of searching for more obscured AGN in dwarf galaxies based on the presence of strong mid-IR fine-structure lines such as $[O\rom{4}](25.89\rm \mu m)$. However, a larger sample is required to reliably establish the connection between $L_{[ O\rom{4}](25.89\rm \mu m)}$ and hard X-ray luminosities  for IMBHs.
  \item The optical to X-ray flux ratio, $\alpha_{\rm OX}$, for J1440 agrees with the empirical relation established between $\alpha_{\rm OX}$ and $L_{2500\text{\AA}}$ for typical type 1 AGNs.  
  This differs from earlier soft X-ray studies of low-mass AGN which typically deviate from this linear relation. Although we do note that measurement of $L_{2500\text{\AA}}$ also has significant uncertainty in it. Similar analysis for other IMBH-powered AGNs with NuSTAR can reveal if this population does follow the extension of the  $\alpha_{\rm OX}-L_{2500\text{\AA}}$ established for more massive AGNs. 
\end{itemize}

This work provides an important step toward understanding the AGN demographics in dwarf galaxies powered by IMBH. We showed that heavy X-ray obscuration can occur in AGNs with IMBHs too.
It is now widely believed that for more massive galaxies, the majority of AGNs are X-ray obscured \citep{hickox2018obscured}. Recently, \citet{ananna2019accretion} using X-ray luminosity functions showed that the Compton-thick fraction in SMBHs could be as high as $\sim50\%$. If IMBHs in dwarf galaxies are assumed to be just an extension of SMBHs in regular galaxies, then we should expect a similarly high fraction of undiscovered Compton-thick AGNs. J1440 provides a unique opportunity to study the realms of those heavily obscured IMBHs in dwarf galaxies. The fact that even medium-deep soft X-ray observations from Chandra and XMM cannot detect the obscured AGN signal buried in the host-galaxy emission highlights the importance of sensitive hard X-ray observations in finding this elusive population. While J1440 is among the highest $\lambda_{\rm Edd}$ dwarf galaxies in the H17 sample, it was still challenging to obtain sufficient hard X-ray counts with a $\approx 100$~ks NuSTAR observation. This highlights the necessity of a next-generation hard X-ray observatory (e.g., HEX-P) for revealing the most elusive AGN population in our cosmic neighborhood. 
Our results also highlight the potential of mid-IR fine-structure lines for discerning AGN signals in systems with heavy host-galaxy contamination. 
As we enter into the age of JWST, a larger sample size of objects similar to J1440 becomes possible for exploring the physical properties of obscured accreting IMBHs. 
\par We would like to thank the anonymous referee for the helpful comments and suggestions which improved the manuscript. We would also like to thank Johannes Buchner for helping us gain familiarity with the BXA software.
This work was supported under NASA grant no. 80NSSC21K0017 and made use of data from the NuSTAR mission, a project led by the California Institute of Technology, managed by the Jet Propulsion Laboratory, and funded by the National Aeronautics and Space Administration. S.A. and C-T.J.Chen also acknowledge support from Chandra X-ray Center AR0-21013A and AR0-21013B. ESK acknowledges financial support from the Centre National d’Etudes Spatiales (CNES).

\software{HEAsoft \citep{2014ascl.soft08004N}, Sherpa \citep{2001SPIE.4477...76F}, CIGALE \citep{boquien2019cigale}, BXA \citep{2014A&A...564A.125B} }
\bibliographystyle{aasjournal}

\end{document}